\documentclass[conference]{IEEEtran}
\IEEEoverridecommandlockouts
\usepackage{cite}
\usepackage{amsmath,amssymb,amsfonts}
\usepackage{algorithmic}
\usepackage{graphicx}
\usepackage{textcomp}
\usepackage[table]{xcolor}
\usepackage{xspace}
\usepackage{multirow}
\usepackage[caption=false]{subfig}

\def\BibTeX{{\rm B\kern-.05em{\sc i\kern-.025em b}\kern-.08em
    T\kern-.1667em\lower.7ex\hbox{E}\kern-.125emX}}

\usepackage[utf8]{inputenc} 
\usepackage[T1]{fontenc}    

\newcommand{\ie}{\emph{i.e.},\xspace}
\newcommand{\eg}{\emph{e.g.},\xspace}
\newcommand{\wrt}{\emph{w.r.t.}\xspace}

\usepackage{amsmath, amsthm, amssymb}

\newtheorem{proposition}{Proposition}

\makeatletter
\newcommand{\linebreakand}{%
\end{@IEEEauthorhalign}
\hfill\mbox{}\par
\mbox{}\hfill\begin{@IEEEauthorhalign}
}
\makeatother

\begin{document}

\title{Layer-refined Graph Convolutional Networks for Recommendation
}

\author{\IEEEauthorblockN{Xin Zhou$^{\dagger}$, Donghui Lin$^{\ddagger}$, Yong Liu$^{\dagger}$\textsuperscript{*}, Chunyan Miao$^{\dagger \natural}$}
	\IEEEauthorblockA{$^{\dagger}$Alibaba-NTU Singapore Joint Research Institute, Nanyang Technological University, Singapore,\\
		$^\ddagger$Faculty of Natural Science and Technology, Okayama University, Okayama, Japan,\\
		$^{\natural}$School of Computer Science and Engineering, Nanyang Technological University, Singapore\\
		xin.zhou@ntu.edu.sg, lindh@okayama-u.ac.jp,  \{stephenliu, ascymiao\}@ntu.edu.sg}
}

\maketitle

\begingroup\renewcommand\thefootnote{*}
\footnotetext{Corresponding Author}

\begin{abstract}
Recommendation models utilizing Graph Convolutional Networks (GCNs) have achieved state-of-the-art performance, as they can integrate both the node information and the topological structure of the user-item interaction graph.
However, these GCN-based recommendation models not only suffer from over-smoothing when stacking too many layers but also bear performance degeneration resulting from the existence of noise in user-item interactions.
In this paper, we first identify a recommendation dilemma of over-smoothing and solution collapsing in current GCN-based models.
Specifically, these models usually aggregate all layer embeddings for node updating and achieve their best recommendation performance within a few layers because of over-smoothing.
Conversely, if we place learnable weights on layer embeddings for node updating, the weight space will always collapse to a fixed point, at which the weighting of the ego layer almost holds all.
We propose a layer-refined GCN model, dubbed LayerGCN, that refines layer representations during information propagation and node updating of GCN. 
Moreover, previous GCN-based recommendation models aggregate all incoming information from neighbors without distinguishing the noise nodes, which deteriorates the recommendation performance.
Our model further prunes the edges of the user-item interaction graph following a degree-sensitive probability instead of the uniform distribution.
Experimental results show that the proposed model outperforms the state-of-the-art models significantly on four public datasets with fast training convergence.
The implementation code of the proposed method is available at https://github.com/enoche/ImRec.
\end{abstract}


\section{Introduction}
\label{sec:introduction}
Recent years have seen a surge in research of graph-based recommendation methods, mainly because the interactive data between users and items are naturally graph-structured.
Nodes in the graph represent either users or items, while edges denote interactions, (\eg clicking, rating, and purchasing), between users and items.
Efficient recommender systems aim to learn both compressed and accurate vectors from these historical interactions to represent users' preferences and items' attributes.
Recent works have shown a boost in performance when introducing graph convolutional networks (GCNs) into recommender systems~\cite{monti2017geometric, wang2019neural, he2020lightgcn}.

The basic intuition behind GCN~\cite{kipf2017semi} is straightforward. \textit{a). Information propagation.} At each iteration, each node uses the first-order local convolution operation to aggregate the information from its local neighborhood.\textit{ b). Node updating.} The node updates its embedding by combining the aggregated information received from neighbors with its current node embedding.
We term the nodes' current embeddings as the ego layer.
With the iteration, the ego layer contains more and more high-order neighborhood information.
In addition, the updated node embedding also encodes the degree information of all the nodes in its high-order neighborhood. 
Hence, the GCN-based recommendation models usually achieve the state-of-the-art recommendation performance.

Although GCN-based models have shown promising results in recommendation and node classification tasks, 
several papers have reported that the models bear unnecessary complexity and redundant computation~\cite{wu2019simplifying, he2020lightgcn}.
Hence, researchers simplify the GCN structure by removing both the non-linear layer and feature transformation layers~\cite{wu2019simplifying, chen2020revisiting, he2020lightgcn}. 
In the simplified recommendation models, \eg LightGCN~\cite{he2020lightgcn}, information is linearly aggregated and propagated via a re-normalized adjacency matrix.
The models ease the computational burden on large user-item interaction graphs but suffer from the over-smoothing problem.
To be precise, the feature propagation smoothens the hidden embeddings locally along the edges of the graph and ultimately encourages similar predictions among locally connected nodes.
Fixed node updating method (\ie mean) in LightGCN makes the over-smoothing problem even worse on dense datasets. The authors of LightGCN place learnable weights to dynamically integrate information from both the ego layer and all hidden layers but find no improvement~\cite{he2020lightgcn}.
We show the problem in Fig.~\ref{fig:intro_prob} with a dense dataset from MOOC platform~\cite{zhang2019hierarchical}.
When we replace learnable weights on layers for dynamic node updating, we find the weighting of the ego layer always dominates others, as shown in Fig.~\ref{fig:intro_prob}. This hampers the information from high-order (\ie several hops away) neighbors integrated into current node representation.
Besides the over-smoothing problem, natural noise can also bias the recommendation due to the uncertainty and vagueness proper of human beings and so on~\cite{yera2016fuzzy}.
Most recommendation models relying on the assumption that the user ratings or implicit feedback can be treated as ground truth of the user's taste may fail to tackle this problem.

\begin{figure}[!tbp]
	\centering
	\includegraphics[trim=10 13 10 10, clip, width=0.45\textwidth]{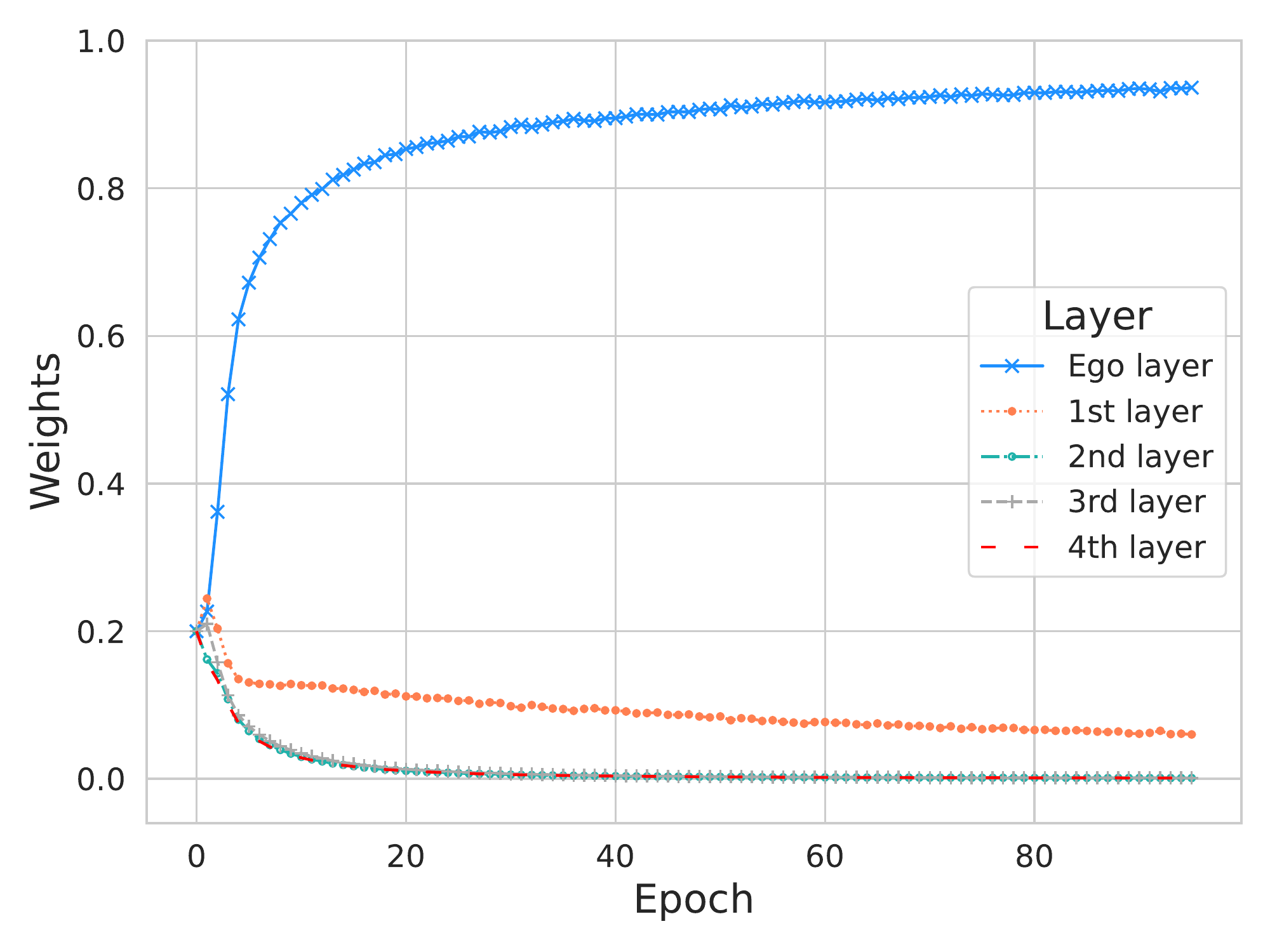}
	\caption{Recommendation dilemma: Fixed node updating results in over-smoothing, on the contrary, the weighting of the ego layer always dominates others if weights on layers are learnable. Experiments conducted with a 4-layer LightGCN~\cite{he2020lightgcn} on the MOOC dataset~\cite{zhang2019hierarchical}.}
	\label{fig:intro_prob}
\end{figure}

A way to address both of these issues is by employing residual and sparsified networks. 
Hence, inspired by ResNet~\cite{he2016deep}, Chen et al. add a residual connection to aggregate embeddings from neighbors~\cite{chen2020revisiting}.
However, the concatenation of embeddings from different layers makes the final node representation hard to focus on informative layers.
Other lines of work address the over-smoothing problem by improving residual connection with initial residual~\cite{chen2020simple} and reversible networks~\cite{li2021training}, etc.
However, these studies are deliberately designed to tackle the node classification problem and should be carefully tuned to be applied in the recommendation domain.
In task perspectives, node classification targets to predict the potential label given the representation of the node. However, recommendation models are required to predict the underlying relation between a user-item pair. That is, whether or not to recommend an item to a target user depends not only on the representation of the user but also on the item.
When considering graph learning, neighbors of a node often share the same label with that node, and their learned representations tend to be similar in node classification. 
Nonetheless, in user-item bipartite graphs, the neighbors of a node (\eg user) belong to a different type (\ie item) and do not necessarily show similar representations.
Stacking too many layers for node classification can include neighbors with different labels into the node's receptive field, which over-smoothes the representation of this node with its high-order neighbors. 
That is why the existing mechanisms that inject the ego layer into high-order graph convolutions for node classification can prevent the nodes' representations from over-smoothing.
However, they use fixed values or manually tuned hyper-parameters to balance the information integrated from the high-order neighbors and their ego representations. Hence, they lack flexibility in adapting to different layers or graphs and may result in suboptimal solutions.
Meanwhile, to mitigate the performance degradation caused by noise interactions, exiting models either drop the interactions randomly~\cite{rong2020dropedge} or learn how to drop~\cite{luo2021learning, schlichtkrull2021interpreting}.
Randomly dropping does not differentiate the importance of edges and hinders the information aggregation of low-degree nodes, resulting in slow training convergence.
The latter methods learn the weights of edges by calculating the attention scores between connected nodes, thus rely heavily on the quality of node representations.

In this paper, we address the over-smoothing problem by refining the node representations in hidden layers of GCN.
The following hidden layers contribute to the final node representation by weighing their similarity with the ego layer. We theoretically prove that our model can alleviate the over-smoothing in existing GCNs.
We further sparsify the user-item graph by pruning its superfluous edges following a degree-sensitive probability distribution, which results in better performance and much faster convergence than random pruning with the uniform distribution.

We summarize the contributions of the paper as follows:
\begin{itemize}
	\item We tackle the over-smoothing issue of GCN-based recommendation models by proposing LayerGCN with hidden layer refinement during network propagation of GCN.
	\item We theoretically demonstrate that LayerGCN is as powerful as the Weisfeiler-Lehman graph isomorphism test~\cite{weisfeiler1968reduction} and can alleviate the over-smoothing problem of existing GCN-based recommendation methods.
	\item To ease the natural noise introduced by external factors in the recommendation scenario, we further sparsify the user-item interaction graph by pruning the edges according to a degree-sensitive probability.
	\item 
	Finally, we conduct comprehensive experiments to show that our proposed model can significantly outperform the state-of-the-art methods with fast training convergence.
\end{itemize}

\section{Related Work}
Recommender systems are proposed to provide users with personalized online products or service recommendations to handle the increasing online information overload problem.
In recommender systems, Collaborative Filtering (CF) is a canonical technique that predicts the interests of a user by aggregating information from similar users.
Here, we first review the classic CF models and then dive into the recent graph neural network-based recommendation models.

\subsection{Classic CF Models}
Classic CF models first find the most similar users or items based on the historical interactions of a target user. 
They then recommend items that similar users have interacted with but not yet interacted by this user.
CF-based recommender systems have many forms.
Based on user-centric or item-centric filtering, they can be divided into user-based CF models and item-based CF models.
Despite the prevalence of CF in modern recommender systems~\cite{covington2016deep, bokde2015matrix}, it has a few limitations, such as scalability and poor performance when processing sparse user-item interaction data.
A more advanced method, Matrix Factorization (MF), decomposes the original sparse user-item interaction matrix into low-dimensional matrices with latent factors/features.
To factorize the matrix, techniques based on dimensional reduction such as Principal Component Analysis (PCA) and Singular Value Decomposition (SVD) are further proposed for building recommender systems~\cite{paterek2007improving, bokde2015matrix}.
The ultimate objective of MF is to learn a set of informative user/item latent features to represent the preferences/attributes of users/items.

\subsection{Graph Neural Networks based Recommendation Models}
With the renaissance of deep learning, neural recommender systems~\cite{wang2015collaborative,he2017neural, zhang2019deep} are proposed to efficiently learn the latent embeddings of users and items.
However, these models fail to capture the structure information of the user-item interaction graph.
Hence, since GCN~\cite{kipf2017semi} was first outlined to generalize convolutional neural networks(CNNs) on graph-structured data, GCN and its variants have been successfully applied to solve various recommendation problems~\cite{berg2018graph,ying2018graph, wang2019neural,he2020lightgcn}.
The basic intuitions are:
1). The available user-item interaction data might be better represented as a bipartite graph;
2). GCN can leverage both information from high-order neighbors and the graph structure.
Especially, GCMC~\cite{berg2018graph} and NGCF~\cite{wang2019neural} are designed under the CF setting.
In these models, GCNs are adopted to propagate the information using the normalized adjacency matrix and aggregate information received from neighbors via the non-linear activation and linear transformation layers.
In~\cite{chen2020revisiting}, a residual preference learning in GCN is designed to tackle the over-smoothing problem and shows good performance.
Although the over-smoothing problem is highly valued and studied in the graph community, it is rarely examined in the recommendation system research domain.
Recent work~\cite{he2020lightgcn} simplifies the vanilla GCN by removing the non-linear activation as well as the feature transformation layer in GCN, which is in line with the research of SGC~\cite{wu2019simplifying}.
Although the over-simplified GCN, LightGCN, is efficient in computation, it is easily stuck into a sub-optimal space with shallow layers, due to the over-smoothing problem.
Recent self-supervised models~\cite{zhang2022diffusion, zhou2021selfcf, lee2021bootstrapping} use graph augmentation to bootstrap the representations of users and items.
However, they also suffer from over-smoothing as they mainly use LightGCN as their backbone network.
To tackle the over-smoothing problem, UltraGCN~\cite{mao2021ultragcn} skips the infinite message passing of graph convolutions by placing a constraint loss on user-item interactions. IMP-GCN~\cite{liu2021interest} performs graph convolutions within a generated subgraph, in which users share similar interests.

Another line of work incorporates side information (\eg item content, knowledge graph) into the CF paradigm. For example, VBPR~\cite{he2016vbpr} extends classic BPR~\cite{rendle2009bpr} by taking into account visual features of items. KGIN~\cite{wang2021learning} encodes item knowledge relations into graph neural networks for item recommendations. Survey papers~\cite{lops2011content} and~\cite{guo2020survey} detail researches and trends on content- and knowledge graph-based recommendations, respectively.

In this paper, we design a novel model (LayerGCN) that refines the hidden layers of a node before updating its representation to tackle the over-smoothing problem.
It is worth noting that although LayerGCN mainly focus on graph learning with nodes' ID embeddings, the proposed model could be applied to other scenarios where nodes are associated with rich semantic features. For example, LayerGCN can be adopted as the backbone network in work of~\cite{zhang2021mining, zhou2022bootstrap, zhou2022tale}. 
Similar to LightGCN~\cite{he2020lightgcn}, LayerGCN can be applied in two different ways to content-based scenarios. First, nodes' initial representations in LayerGCN can be initialized as the content features, which follows GCN~\cite{kipf2017semi} in node classification. Second, ID embeddings generated by LayerGCN can be integrated with content features by advanced operators, such as concatenate, add, attention etc., to obtain the final representations of nodes for recommendation. The work of~\cite{zhang2021mining, zhou2022bootstrap} follows this approach.

\section{Methodology}
\subsection{Preliminaries}
\textbf{Notations.} 
Suppose $\mathcal{G} = (\mathcal{V},\mathcal{E})$ be a given graph with node set $\mathcal{V}$ and edge set $\mathcal{E}$. 
We denote the cardinality of node set $\mathcal{V}$ by $\mathcal{N} = |\mathcal{V}|$ and the cardinality of edge set by $\mathcal{M} = |\mathcal{E}|$. 
The adjacency matrix is denoted by $\mathbf{A} \in \mathbb{R}^{\mathcal{N}\times \mathcal{N}}$, and the diagonal degree matrix is denoted by $\mathbf{D}$.
In the recommendation domain, the user-item interaction graph can be treated by a bipartite graph with source nodes as users and target nodes as items.
The number of users is denoted by $\mathcal{N}_\mathcal{U}$ and the number of items is denoted by $\mathcal{N}_\mathcal{I}$. As a result, we have $\mathcal{N} = \mathcal{N}_\mathcal{U} + \mathcal{N}_\mathcal{I}$.
We denote the embeddings of nodes in the ego layer by $\mathbf{X}^0 \in \mathbb{R}^{\mathcal{N} \times \mathcal{T}}$, where $\mathcal{T}$ is the embedding size of a node. 
In our following discussion, we mainly use the matrix form notation of each model.

\noindent \textbf{GCN.} 
A typical feed forward propagation GCN to calculate the hidden embedding $\mathbf{X}^{l+1}$ at layer $l+1$ is recursively conducted as:
\begin{equation}
	\mathbf{X}^{l+1} = \sigma\left( \hat{\mathbf{A}}\mathbf{X}^{l}\mathbf{W}^{l} \right),
	\label{eq:gcn_va}
\end{equation}
where $\sigma(\cdot)$ is a non-linear function, \eg the ReLu function, $\hat{\mathbf{A}} = \hat{\mathbf{D}}^{-1/2}(\mathbf{A}+\mathbf{I})\hat{\mathbf{D}}^{-1/2}$ is the re-normalization of the adjacency matrix $\mathbf{A}$, and $\hat{\mathbf{D}}$ is the diagonal degree matrix of $\mathbf{A}+\mathbf{I}$.
For node classification, the last layer of a GCN is used to predict the label of a node via a $softmax$ classifier.

\noindent \textbf{LightGCN.} He et al. simplify the vanilla GCN for recommendation in~\cite{he2020lightgcn}.
They find both the feature filter matrix $\mathbf{W}$ and the non-linear activation $\sigma(\cdot)$ impose adverse effects on recommendation performance.
The simplified graph convolutional layer in LightGCN is defined as:
\begin{equation}
	\mathbf{X}^{l+1} = (\mathbf{D}^{-1/2} \mathbf{A} \mathbf{D}^{-1/2})\mathbf{X}^{l}.
	\label{eq:lightGCN_p}
\end{equation}
The node embeddings of the $(l+1)$-th hidden layer are only linearly aggregated from the $l$-th layer with a transition matrix  $\mathbf{D}^{-1/2} \mathbf{A} \mathbf{D}^{-1/2}$.
The transition matrix is exactly the weighted adjacency matrix mentioned above.
Other researchers also verify that the linear transformation matrix of a graph convolutional layer in a GCN only contributes to over-fitting and propose architectures such as SGC~\cite{wu2019simplifying} and DGI~\cite{velickovic2019deep} to solve the over-fitting problem.

Existing GCN-based recommendation models typically integrate both the ego layer and all hidden layers obtained from each propagation to update the final embeddings of nodes. 
Specifically, we use a R{\scriptsize EADOUT} function to aggregate all layer features of a node to obtain the final nodes representation:

\begin{equation}
	\mathbf{X} = \text{R{\scriptsize EADOUT}}(\mathbf{X}^0, \mathbf{X}^1, \mathbf{X}^{2}, \cdots ,\mathbf{X}^{L}),
	\label{eq:lgn_layer_update}
\end{equation}
where the R{\scriptsize EADOUT} function can be any differentiable function and $L$ is the total number of layers in GCN. For example, LightGCN~\cite{he2020lightgcn} uses a mean function for its final embedding updating.

\subsection{LayerGCN}
Note that LightGCN defines the transition matrix as $\mathbf{D}^{-1/2} \mathbf{A} \mathbf{D}^{-1/2}$.
In this paper, we adopt the same transition matrix $\hat{\mathbf{A}} = \mathbf{D}^{-1/2} \mathbf{A} \mathbf{D}^{-1/2}$, which is symmetric. 
The adjacency matrix $\mathbf{A} \in \mathbb{R}^{\mathcal{N} \times \mathcal{N}}$ is constructed from the user-item interaction matrix 
$\mathbf{R} \in \mathbb{R}^{\mathcal{N}_\mathcal{U} \times \mathcal{N}_\mathcal{I}}$:
\begin{equation}
	\mathbf{A} =
	\begin{pmatrix}
	\mathbf{0} & \mathbf{R} \\
	\mathbf{R^T} & \mathbf{0}
	\end{pmatrix},
\end{equation}
and each entry $R_{ui} \in \mathbf{A}$ is set to 1, if user $u$ has interacted with item $i$, otherwise, $R_{ui}$ is set to 0. 

\subsubsection{Degree-sensitive Edge Dropout (DegreeDrop)}
Following the ideas of model sparsification~\cite{rong2020dropedge, louizos2018learning}, we first tackle the natural noise existing in the interaction matrix by pruning graph edges.

In CF, our ultimate objective is to learn a set of informative embeddings for users and items.
Hence, we prune a set of superfluous edges in the original graph following a pre-calculated probability to attenuate the signals from noisy neighbors.
In detail, given a specific edge $e_{k} \in \mathcal{E}, (k < \mathcal{M})$ which connects node $i$ and $j$, its probability to be reserved in graph $\mathcal{G}$ is calculated as:
\begin{equation}
	p_{e_{k}} = \frac{1}{\sqrt {d_i} \sqrt {d_j}},
	\label{eq:degreedrop}
\end{equation}
where $d_i$ and $d_j$ are the degrees of node $i$ and $j$ in graph $\mathcal{G}$, respectively.
With $m$ edges to be pruned (pruning ratio: $m/\mathcal{M} < 1$), we sample $\mathcal{M}-m$ edges from the multinomial probability distribution with parameters $\mathcal{M}-m$ and $\textbf{p}=(p_{e_0},p_{e_1},...,p_{e_{\mathcal{M}-1}})$. Then, we construct a sparsified adjacency matrix $\mathbf{A}_{p}$ based on the sampled edges.
Following~\cite{kipf2017semi, rong2020dropedge}, we also perform the re-normalization trick on $\mathbf{A}_{p}$, resulting as $\hat{\mathbf{A}}_{p}$.
In information propagation of LayerGCN, we will use $\hat{\mathbf{A}}_{p}$ instead of $\mathbf{D}^{-1/2} \mathbf{A} \mathbf{D}^{-1/2}$.
However, in inference stage, we resort to $\mathbf{D}^{-1/2} \mathbf{A} \mathbf{D}^{-1/2}$.
Edge pruning can naturally prevent over-fitting in GCN based on the theoretical study of DropEdge~\cite{rong2020dropedge}.
However, different from DropEdge, which prunes the edges following a uniform distribution and shows slow convergence in model training, the proposed degree-sensitive pruning method tends to drop the edges that connect popular nodes at both ends and converges much faster than DropEdge in practice. It is also verified in~\cite{chen2020simple} that popular nodes are more likely to suffer from over-smoothing.

It is worth noting that edges of the graph can be further permuted with alternating degree-sensitive and random pruning. Alternating edge pruning methods may introduce more diversity into the graph and enjoy the fast training convergence speed.

\subsubsection{Layer-refined Graph Convolution (LayerGC)}
Recent studies~\cite{li2018deeper, oono2020graph, chen2020simple} suggest that GCN does not scale well to deep architectures, since stacking multiple layers of graph convolutions leads to high complexity in back-propagation.
In practice, most GCN variants, including GCN-based recommendation models, achieve the best performance with 2 layers~\cite{kipf2017semi, ying2018graph}.
In CF, our ultimate objective is to learn an informative embedding that represents the personality/property of each user/item.
However, the propagated information via multi-layers tends to be over-smoothed, making it impossible to discriminate between different nodes.
Inspired by the huge success of ResNet~\cite{he2016deep}, recent works~\cite{kipf2017semi, chen2020simple} stimulate the skip connection in ResNet that combines the smoothed node embedding at $(l+1)$-th layer with a fixed weight of information from the $l$-th layer or the ego layer.
Nevertheless, in our propagation stage, we extract the information from the ego layer dynamically by designing a layer refinement mechanism.
Formally, we define the information propagation in LayerGCN as:
\begin{equation}
	\begin{split}
	\mathbf{X}^{l+1} &= \hat{\mathbf{A}}_{p}\mathbf{X}^{l} \\
	\mathbf{X}^{l+1} &= (\mathbf{a}^{l+1} + \epsilon) \mathbf{X}^{l+1},\\
	\end{split}
	\label{eq:propa2}
\end{equation}
where $\mathbf{a}_{l+1} \in \mathbb{R}^\mathcal{N}$ is the node similarity vector between current layer and the ego layer.  $\epsilon$ is a small positive infinitesimal quantity to prevent zero vector in $\mathbf{X}^{l+1}$.That is:
\begin{equation}
	\mathbf{a}^{l+1} = \text{S{\scriptsize IM}}(\mathbf{X}^{l+1}, \mathbf{X}^{0}).
	\label{eq:sim_fun}
\end{equation}
In our model, we use cosine similarity as our S{\scriptsize IM} function based on two principles. First, it is non-parametric. Hence, LayerGCN harnessing this function is less prone to over-fitting. Second, cosine similarity is a widely used metric to evaluate the distance between embeddings and has demonstrated its promising performance in literature~\cite{sundararaman2020methods, zhang2021mining}.
Specifically, the cosine similarity function between two row vectors $\mathbf{x}_i = \mathbf{X}^{l+1}[i,:]$ and $\mathbf{x}_j = \mathbf{X}^{0}[j,:]$ is defined as:
\begin{equation}
	\text{S{\scriptsize IM}}(\mathbf{x}_i, \mathbf{x}_j) = \frac{\mathbf{x}_i \cdot \mathbf{x}_j}{max(||\mathbf{x}_i||_2 ||\mathbf{x}_j||_2, \epsilon)}.
	\label{eq:sim_fun_1}
\end{equation}

This function amplifies the fusion of hidden layers that are similar to the ego layer and reduces the influence of hidden layers that are divergent from the ego layer. 
We will analyze the proposed mechanism which has a variety of merits in the following section in detail.

Finally, we update the nodes' final embeddings by dropping the ego layer as its information is already refined within hidden layers. Hence Eq.~\ref{eq:lgn_layer_update} can be expressed as:
\begin{equation}
	\mathbf{X} =  \text{R{\scriptsize EADOUT}}(\mathbf{X}^1, \mathbf{X}^2, \cdots, \mathbf{X}^L).
	\label{eq:res_update}
\end{equation}
Here we use the sum aggregation for node updating.
In the prediction phase, the final embedding $\mathbf{X}$ is applied to rank the candidate items for each target user.

\begin{figure*}[h]
    \centering
    \includegraphics[trim=25 20 30 0, clip, width=0.96\textwidth]{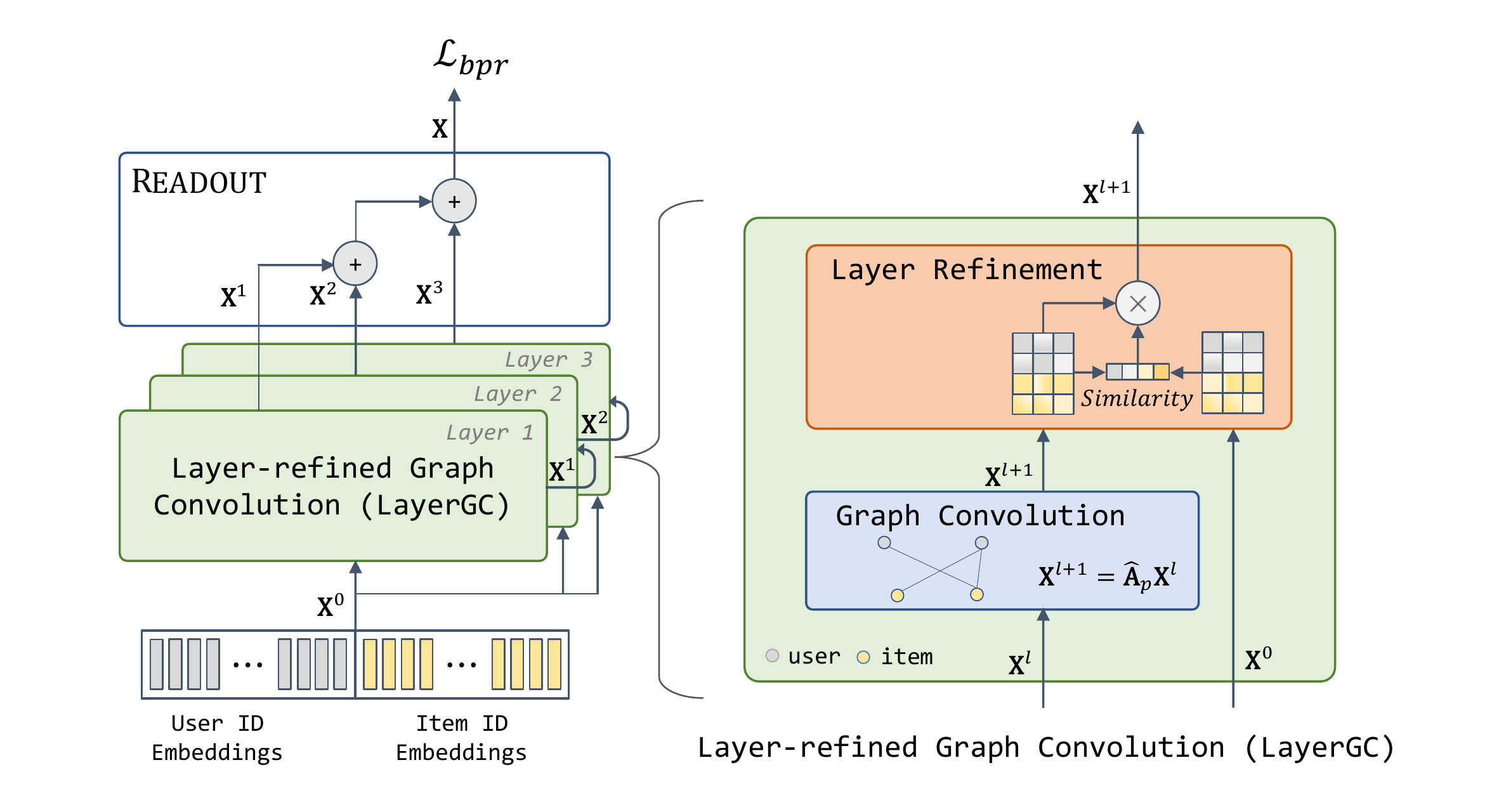}
    \caption{Overview of the proposed LayerGCN. In layer refinement component (\textit{left}), We use the ego layer ($\mathbf{X}^0$) to refine the information of the hidden layer ($\mathbf{X}^l$) before it propagates to the next layer ($\mathbf{X}^{l+1}$).}
    \label{fig:del_gcn}
\end{figure*}

We conclude our proposed model in Fig.~\ref{fig:del_gcn}.
First, to tackle the natural noise in the user-item interaction graph, we sparsify it by pruning a proportion of edges following a degree-sensitive probability.
We then perform linear forward propagation based on the pruned graph.
With each propagated layer embedding, we update its representation based on its similarity with the ego layer.
Finally, we aggregate all embeddings except the ego layer for final node representation.
Based on the final node embedding $\mathbf{X}$, we rank all potential items against each user with the following score for top-$K$ recommendation:
\begin{equation}
	\hat{r}_{ui} = \mathbf{x}_u \mathbf{x}_i^T, \mathbf{x}_u \in \mathbf{X} \cap \mathbf{x}_i \in \mathbf{X},
\end{equation}
where $\mathbf{x}_u \in \mathbb{R}^\mathcal{T}$ and $\mathbf{x}_i \in \mathbb{R}^\mathcal{T}$ are the final embeddings of user $u$ and item $i$, respectively. Note $\mathcal{T}$ is the embedding size of a node (\ie user/item).

\noindent \textbf{The Loss Function.}
We optimize the LayerGCN model under a supervised learning setting.
That is, given a positive user-item pair $(u,i)$ from the interactions, we will sample a negative pair $(u,j)$ that user $u$ has never interacted with item $j$ in the training set.
We adopt the pairwise Bayesian Personalized Ranking (BPR) loss~\cite{rendle2009bpr}, which encourages the prediction of positive user-item pair to be scored higher than its negative pair:
\begin{equation}
	\mathcal{L}_{bpr} = \sum_{(u,i,j)\in \mathcal{D}} -log \sigma(\hat{r}_{ui} - \hat{r}_{uj}),
\end{equation}
where $\mathcal{D}$ is the set of training instances, and each instance $(u,i,j)$ satisfies $r_{u,i} = 1 ~\textrm{and}~ r_{u,j} = 0$.

We further regularize the embeddings of users and items with $L_2$ loss to avoid over-fitting.
The final loss function is defined as:
\begin{equation}
	\mathcal{L} = \mathcal{L}_{bpr} + \lambda \cdot ||\mathbf{X}^0||^2_2,
\end{equation}
where $|| \cdot ||_2$ is $L_2$ norm, $\lambda$ is a penalty term regularizes the coefficients of the model.

\section{Analysis}
We theoretically analyze the merits that can be obtained from the design of LayerGCN and connect our work to the research on residual graph networks.
\subsection{LayerGCN}
We state the proposed LayerGCN enjoys the following properties:
\begin{proposition}
	\label{prop:exp}
	The representational capacity of LayerGCN is as powerful as the Weisfeiler-Lehman (WL) test.
\end{proposition}
The Weisfeiler-Lehman test of graph isomorphism~\cite{weisfeiler1968reduction} is an effective and computationally efficient test that distinguishes a broad class of graphs. 
The algorithm discriminates two graphs as non-isomorphic if at some iteration the embeddings of the nodes between the two graphs differ.
According to Theorem 3 of GIN in~\cite{xu2019powerful}, a GNN architecture is capable of discriminating two different graphs that the Weisfeiler-Lehman test of isomorphism decides as non-isomorphic if its aggregation, updating, and readout functions are injective.
In LayerGCN, both the aggregation and readout functions are the sum aggregator that summarize the information propagated from neighbors. Based on Lemma 5 of~\cite{xu2019powerful}, the sum aggregator can represent injective function over the set of neighborhood embeddings.
It is worth noting the mean aggregator adopted in LightGCN~\cite{he2020lightgcn} is not an injective function, hence is less expressive than LayerGCN.
The layer refinement Eq.~\ref{eq:propa2} is easy to be proved as injective for small embedding values in a range of $[-\pi/4, \pi/4]$.
According to~\cite{kumar2017weight}, the parameter values of deep neural networks in each layer can be captured within this range. 
Furthermore, we can relax $\epsilon$ to $\epsilon >= 1$ to cover any input values while keeping Eq.~\ref{eq:propa2} injective.

\begin{proposition}
	\label{prop:oversmoothing}
	LayerGCN is able to alleviate the over-smoothing of LightGCN.
\end{proposition}
The layer propagation in LightGCN can be denoted as:
\begin{equation}
	\mathbf{X}^{l+1} = \hat{\mathbf{A}}\mathbf{X}^{l}.
	\label{eq:propa3}
\end{equation}
As a result, the $l$-th hidden layer representation of nodes can be derived as:
\begin{equation}
	\mathbf{X}^{l} = \underbrace{\hat{\mathbf{A}} \dotsm \hat{\mathbf{A}}}_{l} \mathbf{X}^0.
	\label{eq:lgn_ll}
\end{equation}
Hence, when the number of layer $l$ goes infinity, we can derive that the representations of nodes pair $(i, j)$ connected by edge $e_{ij} \in \mathcal{E}$ become indistinguishable.
\begin{equation}
	\forall e_{ij} \in \mathcal{E}, \text{we have,} \lim_{l \to \infty} ||\mathbf{x}^l_i - \mathbf{x}^l_j||_p = 0,
	\label{eq:propa4}
\end{equation}
where $||\cdot||_p$ is $L_p$-norm.
The above equation reveals that when stacking with too many layers in LightGCN, the node representations will become indistinguishable and lead to the issue of over-smoothing.

In contrast, our method refines the representation of nodes at $l$-th layer as (we ignore edge pruning and $\epsilon$ for simplicity):

\begin{equation}
	\mathbf{X}^{l} = \text{S{\scriptsize IM}}(\hat{\mathbf{A}} \mathbf{X}^{l-1}, \mathbf{X}^{0}) \mathbf{X}^{l-1}.
	\label{eq:laery_ll}
\end{equation}

To evaluate how the $l$-th layer representation $\mathbf{x}^l$ of a node diverges from its ego representation $\mathbf{x}^0$, we let $p=2$ in Eq.~\ref{eq:propa4} and calculate:
\begin{equation}
	d^l = ||\mathbf{x}^l - \mathbf{x}^0||_2.
	\label{eq:node_layer}
\end{equation}
For analysis simplicity, we assume the hidden representation $\mathbf{x}^l$ propagated at layer $l$ of a node by our model before refinement is the same as that of LightGCN.
Therefore, we can derive the distance $d^l_{lgn}$ between the $l$-th representation and the ego representation of a node is calculated by LightGCN as:
\begin{equation}
	{d^l_{lgn}} = ||\mathbf{x}^l - \mathbf{x}^0||_2.
	\label{eq:dis_lgn}
\end{equation}
Our model calculates the distance after refinement on $\mathbf{x}^l$ is:
\begin{equation}
	{d^l_{lr}} = ||\mathbf{x}^l cos(\mathbf{x}^l, \mathbf{x^0}) - \mathbf{x}^0||_2.
	\label{eq:dis_layer}
\end{equation}
Let $cos(\mathbf{x}^l, \mathbf{x^0})=cos(\theta)$. When $cos(\theta) > 0$, two models integrate similar component from $\mathbf{x}^l$. Here, we quantify the distance of $d_l$ when the two vectors are clearly distinguishable with $cos(\theta) < 0$. We rewrite $(\mathbf{x}^l - \mathbf{x}^0)\cdot (\mathbf{x}^l - \mathbf{x}^0)$ as:
\begin{equation}
	\begin{split}
	(\mathbf{x}^l - \mathbf{x}^0)\cdot (\mathbf{x}^l - \mathbf{x}^0) 
	&= |\mathbf{x}^0|^2 + |\mathbf{x}^l|^2 - 2|\mathbf{x}^l| |\mathbf{x}^0| cos(\theta) \\
	&= |\mathbf{x}^0|^2 + |\mathbf{x}^l| \left(|\mathbf{x}^l| - 2|\mathbf{x}^0| cos(\theta)\right) \\
	&\ge |\mathbf{x}^0|^2 + |\mathbf{x}^l cos(\theta)| \\
	& {\hspace{12pt}} \cdot \left(|\mathbf{x}^l cos(\theta)| - 2|\mathbf{x}^0| cos(\theta)\right) \\
	&= \left(\mathbf{x}^l cos(\theta) - \mathbf{x}^0 \right)\cdot \left(\mathbf{x}^l cos(\theta) - \mathbf{x}^0 \right).
	\end{split}
	\label{eq:dis_lgn_d}
\end{equation}
This suggests $d^l_{lgn} >= d^l_{lr}$, that is, the divergence between of $\mathbf{x}^l$ obtained from LightGCN and $\mathbf{x^0}$ is larger than that of LayerGCN.
In other words, our model can preserve the personalized representation of a node and alleviate the over-smoothing issue that suffered in LightGCN.

Note that the similarity function $cos(x)$ is Lipschitz.
By the Mean Value Theorem, there is a $\eta \in (x,y)$ that satisfies $|cos(x) - cos(y)|=|-sin(\eta)(x - y)|$. Thus, we have:
\begin{equation}
	\begin{split}
		|cos(x) - cos(y)| &= |-sin(\eta)(x - y)|\\
		&\le 1 \cdot |x-y|.
	\end{split}
\end{equation}
The Lipschitz constant is at most 1, so the layer refinement limits the integration of information from hidden layers.
For example, in a corner case, the hidden representation $\mathbf{x}^l$ at layer $l$ and $\mathbf{x}^0$ are perpendicular. Then the refinement function solely integrates $\epsilon \cdot \mathbf{x}^l$ from the $l$-th layer.
This mechanism can prevent the ego presentation of a node from diluting with its neighbors.

\subsection{Relationship with Residual Connections}
We analyze the connections between our model and residual connections in GCN variants~\cite{kipf2017semi, chen2020simple} based on the original graph without pruning.
To focus our analysis, we also simplify the standard GCN with residual connection~\cite{kipf2017semi} and GCNII~\cite{chen2020simple} by removing their feature transformation layer and nonlinear function.
The GCN with a residual connection can be represented as:
\begin{equation}
	\mathbf{X}^{l+1} = \hat{\mathbf{D}}^{-1/2}(\mathbf{A}+\mathbf{I})\hat{\mathbf{D}}^{-1/2} \mathbf{X}^l + \mathbf{X}^l.
\end{equation}
For simplicity, let $\hat{\mathbf{A}} = \hat{\mathbf{D}}^{-1/2}(\mathbf{A}+\mathbf{I})\hat{\mathbf{D}}^{-1/2}$.
Each layer in this GCN variant combines the propagated representation $\hat{\mathbf{A}} \mathbf{X}^l$ and $\mathbf{X}^l$.
We show the variant ultimately puts more weight on $\mathbf{X}^l$ in $\mathbf{X}^{l+1}$ when we re-write the above equation as:
\begin{equation}
	\mathbf{X}^{l+1} = (\hat{\mathbf{A}} + \mathbf{I}) \mathbf{X}^l.
\end{equation}
The self-loop $\mathbf{I}$ is added on the re-normalized adjacency matrix $\hat{\mathbf{A}}$, amplifying the information from its current embedding $\mathbf{X}^{l}$. 
To be precise, we can quantify the magnitude of information received from $u$'s neighbors $\text{N{\footnotesize EIGN}}(u)$ and its current embedding in GCN:
\begin{equation}
	\mathbf{x}_u^{l+1} = \sum_{i\in \text{N{\scriptsize EIGN}}(u)\cup \{u\}} \frac{1}{\sqrt{d_{u}}\sqrt{d_{i}}} \mathbf{x}_i^l.
	\label{eq:gcn_node_pro}
\end{equation}
With residual connections in GCN, the above equation can be updated as:
\begin{equation}
	\mathbf{x}_u^{l+1} = \mathbf{x}_u^l + \sum_{i\in \mathrm{NEIGN}(u)\cup \{u\}} \frac{1}{\sqrt{d_{u}}\sqrt{d_{i}}} \mathbf{x}_i^l.
	\label{eq:gcn_node_pro2}
\end{equation}
Compared Eq.~\ref{eq:gcn_node_pro2} with Eq.~\ref{eq:gcn_node_pro}, we can observe the signal from $\mathbf{x}_u^l$ is amplified by $d_u$ in $\mathbf{x}_u^{l+1}$.

Experimental results in~\cite{kipf2017semi} show that the performance on node classification degrades as we stack more layers, but still better than the vanilla version. Because the residual version is capable of preserving the node's ego embedding and preventing it from over-smoothing.
Following this, Chen et al. propose GCNII~\cite{chen2020simple} that constructs a connection to the ego layer $\mathbf{X}^0$ directly: 
	$\mathbf{X}^{l+1} = (1-\alpha)\hat{\mathbf{A}} \mathbf{X}^l + \alpha \mathbf{X}^0,$
where $\alpha$ is a hyper-parameter that controls the information from the initial residual connection $\mathbf{X}^0$.
The hyper-parameter $\alpha$ is fixed once set and usually kept at a low ratio to ensure new information from high-order neighbors can contribute more in $\mathbf{X}^{l+1}$.
From the discussion, we conclude that the GCN variants seek to balance the information aggregation of nodes between high-order neighbors and their current embeddings.
It is also worth noting that both the residual version of GCN and GCNII are designed to prevent the node ego representation from diluting with propagated neighbor information.
That is why adding residual connections from either the previous layer~\cite{kipf2017semi} or the initial layer~\cite{chen2020simple} shows better performance.
However, both adopt fixed value or hyper-parameter to control the ratio of integration of propagated information.
In LayerGCN, we use layer refinement to distill the received information before integrating into the node representations.
We dynamically weigh and update the hidden layer embeddings by calculating their similarities with the ego layer.

\subsection{Computational Complexity}
The computational cost of LayerGCN mainly occurs in linear propagation of $\hat{\mathbf{A}}$.
The analytical complexity of LightGCN is $\mathcal{O}(2L\mathcal{M}\mathcal{T}/\mathcal{B})$, where $L$ is the number of GCN layers and $\mathcal{B}$ is the training batch size.
In graph convolutional operation, our model further adds a similarity function to refine the information in each propagated layer.
The cost for layer refinement is $\mathcal{O}(L\mathcal{N}\mathcal{T}/\mathcal{B})$, where $\mathcal{N}$ is the number of nodes and usually is an order of magnitude smaller than the number of edges $\mathcal{M}$ in graph $\mathcal{G}$. Hence, the computational complexity of LightGCN and
LayerGCN are within the same magnitude.

It is worth noting that LayerGCN is much more computationally efficient than the graph attention mechanism~\cite{velivckovic2017graph, brody2022how}.
In GATs~\cite{velivckovic2017graph, brody2022how}, 
the researchers update a node's representation by injecting learnable weights on neighboring nodes to differentiate the importance of neighbors.
In LayerGCN, we first integrate information from neighboring nodes linearly to obtain a temporal representation of a node.
Then we refine the representation by weighing its similarity with the ego embedding. 
Our model introduces no extra learnable parameters and thus can prevent the model from over-fitting and is more efficient in computation.

\section{Experiments}
We evaluate the performance of LayerGCN as well as the DegreeDrop mechanism on four public datasets.

\subsection{Experimental Settings}
For a fair comparison, we implement all models under a unified framework.
In this framework, we follow the chronological data splitting strategy~\cite{ji2020critical}. Specifically, we first sort all the observed user-item interactions of each dataset in chronological order based on the interaction timestamps. Then, the first 70\% user-item interactions are used as training data, the next 10\% interactions are used as validation data, and the remaining 20\% interactions are used as testing data. In the validation and testing data, we remove the cold-start users and items that have not occurred in the training data. Moreover, we treat each observed user-item interaction in the training data as a positive instance and randomly sample its negative counterpart.

\begin{table}
	\caption{Statistics of the experimented datasets.}
	\centering	
	\def\arraystretch{1.0}
	\setlength\tabcolsep{5.0pt}
	\begin{tabular}{l|r|r|r|r}
		\hline
		Datasets & \# Users & \# Items & \# Interactions & Sparsity \\
		\hline
		MOOC & 82,535 & 1,302 & 458,453 & 99.5734\% \\
        Games & 50,677 & 16,897 & 454,529 &  99.9469\%\\
        Food & 115,144 & 39,688 & 1,025,169 &  99.9776\%\\
		Yelp & 99,010 & 56,441 & 2,762,088 & 99.9506\%\\
		\hline
	\end{tabular}
	\label{tab:dataset}
\end{table}

\subsubsection{Dataset Description}
We consider the diversity of evaluated datasets on the aspects of sparsity, number of users/items, domain or genre, release date.
Based on the consideration, we select the following four public datasets.

\begin{itemize}
	\item MOOC \footnote{Dataset can be accessed at: http://moocdata.cn}: This educational dataset records the enrollment courses of users ranging from October 1st, 2016 to March 31st, 2018~\cite{zhang2019hierarchical}.
	The dataset shows a typical recommendation pattern in a start-up platform, where the number of users is tens or hundreds of the available items.
	\item Amazon Video Games (denoted by ``Games'')~\footnote{Both categories of Amazon review datasets can be downloaded at: https://nijianmo.github.io/amazon}: This dataset and the Grocery and Gourmet Food dataset are subsets of the newly released Amazon review dataset in 2018~\cite{ni2019justifying}.
	The new version has much larger interactions and graph sizes.
	Each entry in the dataset records the rating value of a user-item pair.
	Similar to~\cite{he2016fast, tan2020learning},we treat each rating record as a positive user-item interaction pair.
	\item Amazon Grocery and Gourmet Food (denoted by ``Food''): This dataset is similar to the Amazon Video Games dataset but in the genre of food. Compared with Games dataset, it has a much larger interaction graph.
	\item Yelp \footnote{Dataset can be accessed at: https://www.yelp.com/dataset}: This is a popularly used check-in dataset for business recommendation. The previous version of 2018 lacking time information, so we use the latest version downloaded directly from Yelp.com. 
\end{itemize}

Similar to~\cite{ni2019justifying}, both Games and Food datasets are preprocessed with a 5-core setting on both items and users.
Following~\cite{he2020lightgcn}, we preprocess the Yelp dataset with a 10-core setting on both items and users.
The statistics of datasets are summarized in Table~\ref{tab:dataset}.

\begin{table*}[!htp]
	\centering
	\def\arraystretch{1.0}
	\setlength\tabcolsep{3.2pt} 
	\caption{Overall performance comparison. We mark the global best results on each dataset under each metric in \textbf{bold} face. As the variant of LayerGCN, LayerGCN (w/o Dropout) almost outperforms all the baselines on every dataset under every metric, we mark the best performance from baselines \underline{underlined}. }
	\begin{tabular}{ll|ccccccccc|ccr}
		\hline
		\multirow{2}{*}{Datasets} & \multirow{2}{*}{Metrics} & \multirow{2}{*}{BPR} & \multirow{2}{*}{MultiVAE} & \multirow{2}{*}{EHCF} & \multirow{2}{*}{BUIR} & \multirow{2}{*}{NGCF} & \multirow{2}{*}{LR-GCCF} & \multirow{2}{*}{LightGCN}  & \multirow{2}{*}{UltraGCN} & \multirow{2}{*}{IMP-GCN} & LayerGCN & LayerGCN & $improv.$\\
		& & & &  & & & & & & & (w/o Dropout) & (Full) & (\%)\\
		\hline
		\multirow{6}{*}{MOOC} & R@10 & 0.2424 & \underline{0.2680} & 0.2366 & 0.2354 & 0.2486 & 0.2466 & 0.2507 & 0.2481 & 0.2439 & 0.2615 & \textbf{0.2801*} & 4.51 \\
		& R@20 & 0.3479 & \underline{0.3639} & 0.3307 & 0.3196 & 0.3426 & 0.3336 & 0.3321 & 0.3183 & 0.3345 &  0.3889 & \textbf{0.3979*} & 9.34 \\
		& R@50 & 0.4963 & \underline{0.5373} & 0.4535 & 0.4801 & 0.4943 & 0.4809 & 0.4844 & 0.4687 & 0.4985 & 0.5420 & \textbf{0.5572*} & 3.70 \\
		& N@10 & 0.1576 & 0.1773 & 0.1562 & 0.1589 & \underline{0.1797} & 0.1678 & 0.1588 & 0.1758 & 0.1624 & 0.1754 & \textbf{0.1909*} & 6.23 \\
		& N@20 & 0.1900 & 0.2055 & 0.1841 & 0.1835 & \underline{0.2079} & 0.1938 & 0.1835 & 0.1969 & 0.1894 & 0.2154 & \textbf{0.2272*} & 10.56 \\
		& N@50 & 0.2265 & \underline{0.2476} & 0.2135 & 0.2224 & 0.2443 & 0.2294 & 0.2208 & 0.2324 & 0.2296 & 0.2526 & \textbf{0.2665*} & 7.63 \\
		\hline
		\multirow{6}{*}{Games} & R@10 & 0.0210 & 0.0238 & 0.0278 & 0.0227 & 0.0254 & 0.0259 & 0.0275 & \underline{0.0290} & 0.0277 & \textbf{0.0298} & 0.0296 & 2.07 \\
		& R@20 & 0.0369 & 0.0376 & 0.0445 & 0.0384 & 0.0425 & 0.0446 & 0.0461 & \underline{0.0480} & 0.0477 & 0.0490 & \textbf{0.0502*} & 4.58 \\
		& R@50 & 0.0699 & 0.0718 & 0.0772 & 0.0749 & 0.0825 & 0.0824 & 0.0841 & \underline{0.0879} & 0.0853 & 0.0886 & \textbf{0.0917*} & 4.32 \\
		& N@10 & 0.0135 & 0.0154 & 0.0175 & 0.0143 & 0.0166 & 0.0171 & 0.0175 & \underline{0.0191} & 0.0184 & 0.0185 & \textbf{0.0192} & 0.52 \\
		& N@20 & 0.0183 & 0.0196 & 0.0227 & 0.0192 & 0.0217 & 0.0228 & 0.0231 & \underline{0.0250} & 0.0245 & 0.0243 & \textbf{0.0254} & 1.60 \\
		& N@50 & 0.0265 & 0.0280 & 0.0308 & 0.0282 & 0.0314 & 0.0320 & 0.0326 & \underline{0.0348} & 0.0337 & 0.0340 & \textbf{0.0355} & 2.01 \\
		\hline
		\multirow{6}{*}{Food} & R@10 & 0.0138 & 0.0133 & 0.0158 & 0.0145 & 0.0158 & 0.0172 & 0.0184 & \underline{0.0187} & 0.0184 & 0.0195 & \textbf{0.0199*} & 6.42 \\
		& R@20 & 0.0222 & 0.0208 & 0.0243 & 0.0236 & 0.0254 & 0.0277 & 0.0286 & 0.0287 & \underline{0.0288} & 0.0303 & \textbf{0.0318*} & 10.42 \\
		& R@50 & 0.0390 & 0.0374 & 0.0416 & 0.0469 & 0.0456 & 0.0478 & 0.0497 & 0.0501 & \underline{0.0507} & 0.0525 & \textbf{0.0540*} & 6.51 \\
		& N@10 & 0.0097 & 0.0092 & 0.0111 & 0.0111 & 0.0102 & 0.0120 & 0.0125 & \underline{0.0129} & 0.0127 & 0.0131 & \textbf{0.0139*} & 7.75 \\
		& N@20 & 0.0124 & 0.0116 & 0.0137 & 0.0141 & 0.0132 & 0.0154 & 0.0157 & \underline{0.0161} & 0.0160 & 0.0166 & \textbf{0.0177*} & 9.94 \\
		& N@50 & 0.0167 & 0.0159 & 0.0182 & 0.0201 & 0.0185 & 0.0206 & 0.0211 & \underline{0.0216} & \underline{0.0216} & 0.0223 & \textbf{0.0234*} & 8.33 \\
		\hline
		\multirow{6}{*}{Yelp} & R@10 & 0.0224 & 0.0244 & 0.0275 & 0.0251 & 0.0235 & 0.0259 & 0.0291 & \underline{0.0300} & 0.0297 & 0.0313 & \textbf{0.0319*} & 6.33 \\
		& R@20 & 0.0393 & 0.0423 & 0.0464 & 0.0421 & 0.0413 & 0.0439 & 0.0501 & \underline{0.0508} & 0.0501 & 0.0531 & \textbf{0.0542*} & 6.69 \\
		& R@50 & 0.0799 & 0.0839 & 0.0885 & 0.0824 & 0.0830 & 0.0862 & 0.0967 & \underline{0.0968} & 0.0965 & 0.1004 & \textbf{0.1024*} & 5.79 \\
		& N@10 & 0.0207 & 0.0221 & 0.0260 & 0.0243 & 0.0220 & 0.0239 & 0.0275 & \underline{0.0279} & 0.0276 & 0.0288 & \textbf{0.0294*} & 5.38 \\
		& N@20 & 0.0260 & 0.0275 & 0.0315 & 0.0292 & 0.0275 & 0.0292 & 0.0337 & \underline{0.0341} &  0.0337 & 0.0353 & \textbf{0.0360*} & 5.57 \\
		& N@50 & 0.0380 & 0.0398 & 0.0438 & 0.0410 & 0.0398 & 0.0414 & 0.0473 & \underline{0.0477} & 0.0473 & 0.0491 & \textbf{0.0501*} & 5.03 \\
		\hline		
		\multicolumn{14}{l}{\begin{tabular}{@{}l@{}l@{}}* To ensure that the performance of LayerGCN is not just a consequence of the random seed, \ie to verify the stability of our method, we conduct \\ ~~ experiments of LayerGCN (Full) against the best baseline across 5 different seeds and state the reported performance marked with `*' is significant \\ ~~~at the level of $p < 0.05$ with a paired $t$-test. \end{tabular}}%
\end{tabular}
\label{tab:performance}
\end{table*}

\subsubsection{Baseline Methods}
We compare several representatives and state-of-the-art models, ranging from traditional matrix factorization methods to recent GCN-based models. 
\begin{itemize}
	\item \textbf{BPR}~\cite{rendle2009bpr}: This is a matrix factorization model optimized by a pairwise ranking loss in a Bayesian way.
	\item \textbf{MultiVAE}~\cite{liang2018variational}: It is a generative model that adopts variational auto-encoder (VAE) for item-based CF.
	\item \textbf{EHCF}~\cite{chen2020efficient}: This is an efficient recommendation model that reconstructs the interaction matrix without negative sampling.
	\item \textbf{BUIR}~\cite{lee2021bootstrapping}: This framework uses asymmetric network architecture to update its backbone network parameters. It learns the representations of users and items without using negative samples. We use the BUIR framework encapsulated with LightGCN as its backbone network because of its better performance as revealed in~\cite{lee2021bootstrapping}.
	\item \textbf{NGCF}~\cite{wang2019neural}: This model explicitly injects collaborative signal from high-order connectivity of user-item graph into the embedding learning process.
	\item \textbf{LR-GCCF}~\cite{chen2020revisiting}: The model first simplifies the vanilla GCN by removing the nonlinear function, then it uses a residual preference learning process for prediction.
	\item \textbf{LightGCN}~\cite{he2020lightgcn}: This is a simplified graph convolution network that only performs linear propagation and aggregation between neighbors.
	The hidden layer embeddings are averaged to calculate the final user/item embeddings for prediction.
	\item \textbf{UltraGCN}~\cite{mao2021ultragcn}: This model further simplifies LightGCN by removing the stacking of many graph convolution layers in GCN. It approximates the limit of infinite-layer graph convolutions by placing constraint coefficients on user-item interaction pairs, which is termed constraint loss in the paper.
	\item \textbf{IMP-GCN}~\cite{liu2021interest}: This model splits users into subgroups based on their interests and performs high-order graph convolution within each subgroup.
\end{itemize}
To better understand the behavior of LayerGCN, we add a variant of our model that removes the degree-sensitive edge dropout method for comparison, denoted as LayerGCN (w/o Dropout). The LayerGCN with DegreeDrop mechanism is denoted as LayerGCN (Full).

\subsubsection{Evaluation Metrics}
We use Recall@K and NDCG@K computed by the all-ranking protocol as the evaluation metrics for recommendation performance comparison.
In the recommendation phase, all items that have not interacted with a specific user are regarded as candidates.

Formally, we define $I^r_u(i)$ as the $i$-th ranked item recommended for $u$, $\mathbb{I}[\cdot]$ is the indicator function, and $I^t_u$ is the set of items that user $u$ interacted in the testing data.
Recall@K for an individual user $u$ is defined as:
\begin{equation}
	Recall@K(u) = \frac{\sum^K_{i=1} \mathbb{I}[I^r_u(i) \in I^t_u]}{|I^t_u|}.
\end{equation}

The Discounted Cumulative Gain (DCG@K) for the user $u$ is defined as:
\begin{equation}
	DCG@K(u) = \sum^K_{i=1} \frac{2^{\mathbb{I}[I^r_u(i) \in I^t_u]}-1}{log(i+1)}.
\end{equation}
$NDCG@K$ is normalized to [0, 1] with $NDCG@K=DCG@K/IDCG@K$, where $IDCG@K$ is calculated by sorting the interacted items in the testing data at top $K$.
We set $K=10$, $K=20$ and $K=50$ in our experimental comparison.
For simplicity and formatting considerations, we denote Recall@K as R@K and NDCG@K as N@K in the following sections.

\subsubsection{Hyper-parameter Settings}
Same as other works~\cite{chen2020revisiting, he2020lightgcn}, we fix the embedding size of both users and items to 64 for all models, initialize the embedding parameters with the Xavier method~\cite{glorot2010understanding}, and use Adam~\cite{kingma2015adam} as the optimizer.
For a fair comparison, we carefully tune the hyper-parameters of each model following their published papers. 
If the hyper-parameters are not specified, we seek their ranges in the ablation studies of the papers.
For LayerGCN, we fix its number of layers to 4, tune the coefficient $\lambda$ of $L_2$ regularization term in \{$1e^{-2}, 1e^{-3}, 1e^{-4}, 1e^{-5}$\}, and tune the degree-sensitive edge dropout ratio in $\{0.0, 0.1, 0.2\}$.
The edge dropout ratio of 0.0 means we use the LayerGCN without edge pruning.
For converge consideration, the early stopping and total epochs are fixed at 50 and 1000, respectively.
We implement the proposed model using  PyTorch~\cite{NEURIPS2019_bdbca288}.
All models are trained on a single GeForce RTX 2080 Ti with 12 GB memory.

\subsection{Performance Comparison}
Table~\ref{tab:performance} shows the overall performance comparison with baselines.
Improvement percentage ($improv.$) is calculated as $(i-j)*100/j$, where $i$ and $j$ are the performance of LayerGCN (Full) and the best baseline under each dataset and metric, respectively. 
With BPR as the baseline, we observe that EHCF without negative sampling performs better in reconstructing the original user-item interaction matrix.
Its performance is competitive with the recent graph-based models (\ie BUIR, NGCF, LR-GCCF, LightGCN, UltraGCN and IMP-GCN).
MultiVAE leveraging variational autoencoder has better performance on the dense dataset (\ie MOOC). However, its performance is degraded on sparse datasets (\ie Games and Food).
In general, graph-based models show better performance than others.
The results validate that the simplified GCN without non-linear activation benefits recommendation.
LightGCN further improves over LR-GCCF, which proves it is worth removing the linear transformation $\mathbf{W}$ in Eq.~\ref{eq:gcn_va}~\cite{he2020lightgcn}.
The performance of UltraGCN and IMP-GCN is superior to LightGCN, which demonstrates they can partially alleviate the over-smoothing problem. With the constraint loss and an additional item-item graph in UltraGCN, it outperforms LightGCN on ranking metrics across all datasets. However, we observe the performance of all graph-based CF models degrades under the dense MOOC dataset.
The performance of LayerGCN without edge dropout shows competitive results compared with the full version of LayerGCN on all datasets as it also drops the ego layer in GCN. 
The full version of LayerGCN with edge dropout significantly improves its variant as it further sparsifies the interaction graph.
In addition, our model can fuse more relevant information from neighbors for final node updating.

From the analysis and experimental results, we conclude the behavior of LayerGCN as: \textit{a).} LayerGCN drops the ego layer and distills the propagated information with the layer refinement enabling the model to integrate more signals from neighbors. \textit{b).} The degree-sensitive edge pruning can filter out irrelevant signals in neighbors and achieve better performance.

We further report that LightGCN achieves its best performance on MOOC within 3-layer GCN, as shown in Table~\ref{tab:layer-lgn}. 
For space consideration, we only report $K=20$ and $K=50$ to fit the table into a single column.
In Table~\ref{tab:layer-lgn}, when stacking LightGCN with 4 layers, due to the problem of over-smoothing, its performance on all metrics except Recall@20 degrades.
Conversely, LayerGCN can extract the informative message from the ego layer in layer propagation via the layer refinement mechanism. As a result, it shows higher performance even though its layers are fixed at 4. 

Finally, it is worth noting that the performance of LayerGCN under all datasets are evaluated by fixing the number of GCN layers to 4, while LightGCN searches its best performance regarding GCN layers within the range of [1, 4].

\subsection{Efficiency and Effectiveness of DegreeDrop}
In previous sections, we conducted experiments to verify the proposed model stacking fixed layers can outperform LightGCN.
Here, we further examine the improved edge dropout algorithm, DegreeDrop, with regard to \textbf{convergence} and \textbf{performance}.

\begin{table}
	\centering	
	\def\arraystretch{1.0}
	\setlength\tabcolsep{8.0pt}
	\caption{Comparison of recommendation accuracy on LayerGCN and LightGCN \wrt different layers on the MOOC dataset. LightGCN uses layers under a small number to tackle over-smoothing (Refer. Fig.~\ref{fig:layer-mooc}).}
	\begin{tabular}{c|c|c|c|l}
		\hline
		Model & R@20 & R@50 & N@20 & N@50 \\
		\hline
		LayerGCN - 4 Layers & 0.3979 & 0.5572 & 0.2272 & 0.2665 \\
		\hline
		LightGCN - 4 Layers & 0.3321 & 0.4844 & 0.1835 &  0.2208 \\
		LightGCN - 3 Layers & 0.3271 & 0.4921 & 0.1929 &  0.2330 \\
		LightGCN - 2 Layers & 0.3286 & 0.5140 & 0.1853 & 0.2315 \\
		LightGCN - 1 Layers & 0.3227 & 0.4826 & 0.1844 & 0.2237 \\
		\hline
	\end{tabular}
	\label{tab:layer-lgn}
\end{table}

\subsubsection{Convergence: DegreeDrop \textit{vs.} DropEdge~\cite{rong2020dropedge}} 
We compare LayerGCN using both DegreeDrop and DropEdge under different dropout ratios ranging from 0.1 to 0.8 with a step of 0.1.
For convergence consideration, we record the best epoch as the epoch that achieves the best validation score.
We plot the comparison of DegreeDrop and DropEdge regarding convergence on the MOOC dataset in Fig.~\ref{fig:perform_convergence}(a).
From the figure, we observe DegreeDrop converges faster than DropEdge under every dropout ratio. 
DropEdge removes a certain number of edges from the input graph in a random manner without considering the importance of the edges.
On the contrary, DegreeDrop prunes the superfluous edges according to the high degrees of nodes.
The training epochs required for convergence are only half of DropEdge under a higher edge dropout ratio (\ie $>$ 0.5). 
When considering all dropout ratios, the proposed DegreeDrop can reduce the training epochs of DropEdge by 39\%.

We further plot the batch loss within each epoch for DegreeDrop and DropEdge in Fig.~\ref{fig:perform_convergence}(b) under edge dropout ratio of 0.7.
The curve shows that the batch loss with DegreeDrop descends much faster than DropEdge from the very beginning.

\begin{figure}
	\centering
	\subfloat[Comparison of convergence between EdgeDrop~\cite{rong2020dropedge} and DegreeDrop under different edge dropout ratio on the MOOC. EdgeDrop randomly drops the edges in a graph. Lower best epoch index converges faster.]{\includegraphics[width=0.47\linewidth]{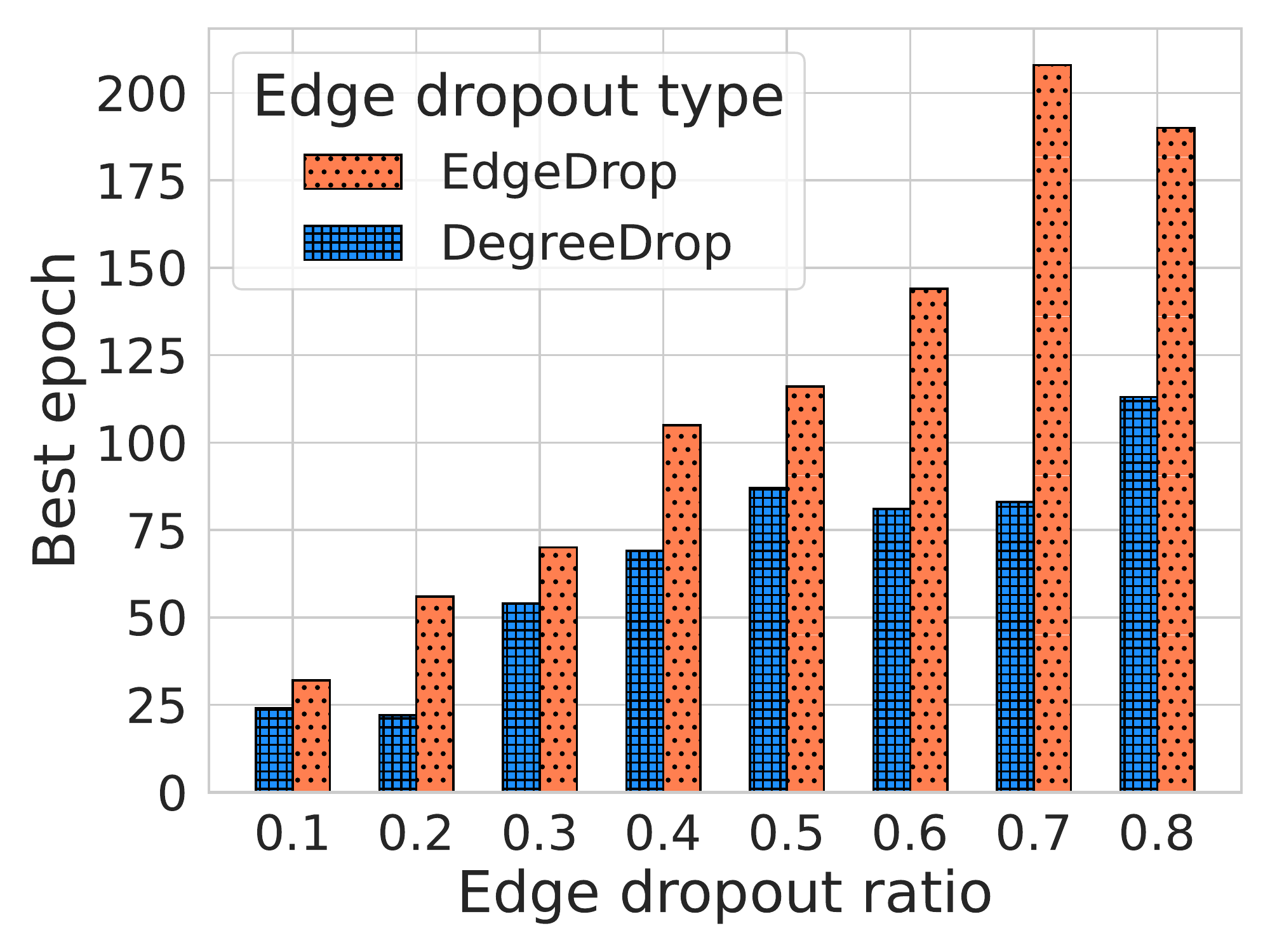} \label{fig:convergence}} \hspace{0.1cm}
	\subfloat[Detailed comparison of convergence curves on batch loss under dropout ratio of 0.7. Lower loss value converges faster.]{\includegraphics[width=0.47\linewidth]{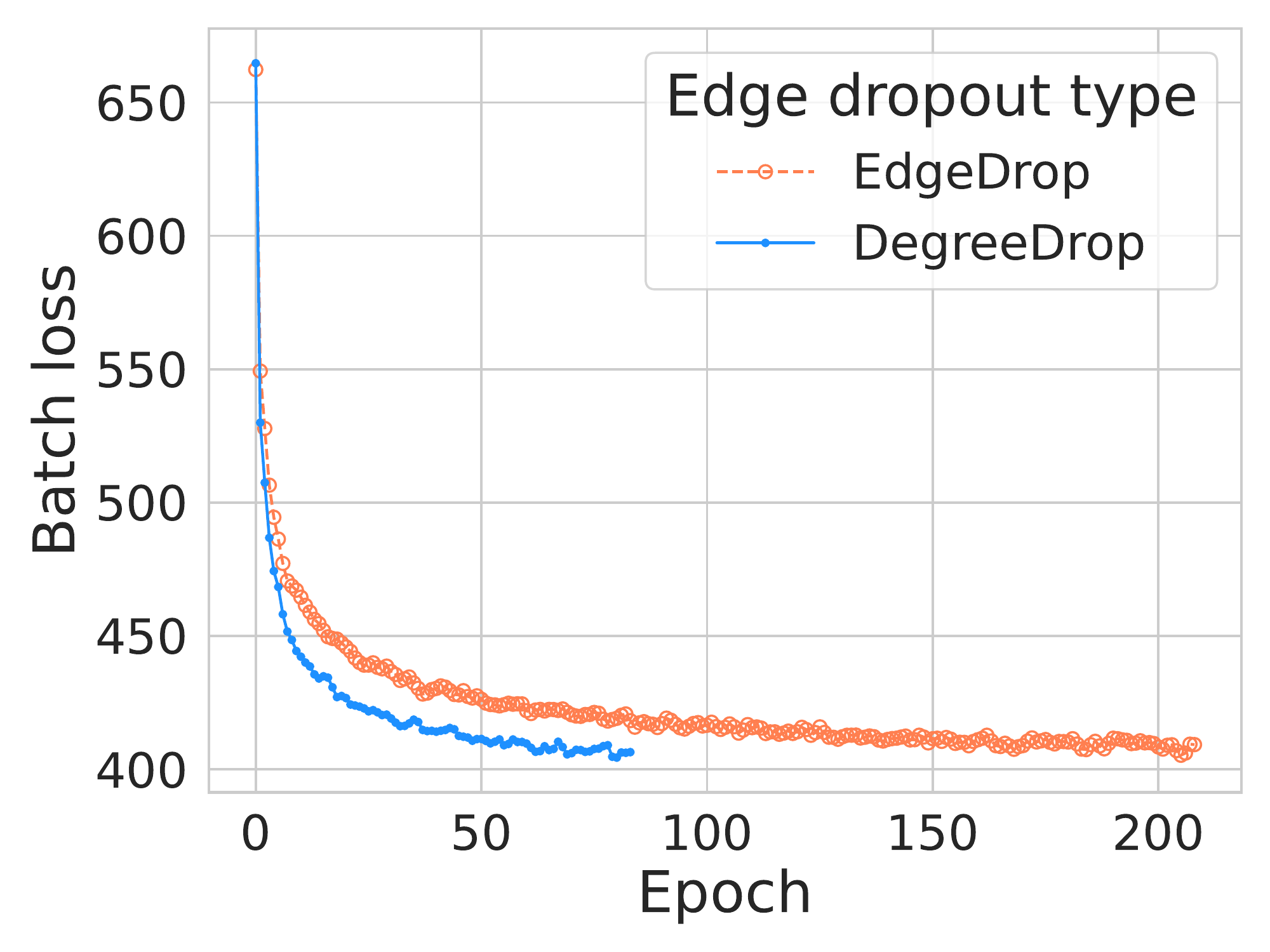} \label{fig:loss_curv}}
	\caption{Convergence rate compared with different edge dropout types of LayerGCN.}
	\label{fig:perform_convergence}
\end{figure}

\begin{table}[!tbp]	
	\def\arraystretch{1.0}
	\setlength\tabcolsep{6.0pt}
	\centering
	\caption{Comparison of recommendation accuracy between DegreeDrop \textit{vs.} DropEdge~\cite{rong2020dropedge} on four datasets with regard to different training epochs. The results show LayerGCN with DegreeDrop achieves better performance under both the same and best epochs.}
	\begin{tabular}{l|l|l|cccc}
		\hline
		Datasets & Variants & Epoch & R@20 & R@50 & N@20 & N@50\\
		\hline %
		\multirow{6}{*}{MOOC} & DropEdge & 20 & 0.3526 & 0.5432 & 0.1964 & 0.2427 \\
		& DegreeDrop & 20 & 0.3637 & 0.5583 & 0.2079 & 0.2579 \\	
		& DropEdge & 50 & 0.3571 & 0.5510 & 0.1940 & 0.2421 \\
		& DegreeDrop & 50 & 0.3979 & 0.5572 & 0.2272 & 0.2665 \\
		& DropEdge & Best & 0.3571 & 0.5510 & 0.1940 & 0.2421 \\
		& DegreeDrop & Best & 0.3979 & 0.5572 & 0.2272 & 0.2665 \\
		\hline
		\multirow{6}{*}{Games} & DropEdge & 20 & 0.0453 & 0.0861 & 0.0218  & 0.0320 \\
		& DegreeDrop & 20 & 0.0463 & 0.0870  & 0.0225 & 0.0325 \\		
		& DropEdge & 50 & 0.0489 & 0.0882 & 0.0234 & 0.0332 \\
		& DegreeDrop & 50 & 0.0485 & 0.0888 & 0.0243 & 0.0341  \\
		& DropEdge & Best & 0.0489 & 0.0882 & 0.0234 & 0.0332 \\
		& DegreeDrop & Best & 0.0502 & 0.0917 & 0.0254 & 0.0355 \\
		\hline
		\multirow{6}{*}{Food} & DropEdge & 20 & 0.0280 & 0.0511 & 0.0153  & 0.0213 \\
		& DegreeDrop & 20 & 0.0283 & 0.0513 & 0.0162 & 0.0221 \\		
		& DropEdge & 50 & 0.0309 & 0.0539 & 0.0170 & 0.0229 \\
		& DegreeDrop & 50 & 0.0315 & 0.0546 & 0.0175 & 0.0235 \\
		& DropEdge & Best & 0.0309 & 0.0530 & 0.0171 & 0.0228 \\
		& DegreeDrop & Best & 0.0318 & 0.0540 & 0.0177 & 0.0234 \\
		\hline
		\multirow{6}{*}{Yelp} & DropEdge & 20 & 0.0377 & 0.0743 & 0.0251  & 0.0360 \\
		& DegreeDrop & 20 & 0.0388 & 0.0767 & 0.0262 & 0.0376 \\		
		& DropEdge & 50 & 0.0429 & 0.0849 & 0.0291 & 0.0415 \\
		& DegreeDrop & 50 & 0.0437 & 0.0864 & 0.0295 & 0.0422 \\
		& DropEdge & Best & 0.0532 & 0.1017 & 0.0350 & 0.0493 \\
		& DegreeDrop & Best & 0.0542 & 0.1024 & 0.0360 & 0.0501 \\
		\hline
	\end{tabular}
	\label{tab:full_perform}
\end{table}

\subsubsection{Performance: DegreeDrop \textit{vs.} DropEdge~\cite{rong2020dropedge}}

For both performance and convergence consideration, we sample epochs of 20 and 50 on all test datasets in Table~\ref{tab:full_perform}.
On four datasets, LayerGCN harnessed with DegreeDrop outperforms that of DropEdge under every metric on the best epoch.
From Table~\ref{tab:full_perform}, we can also observe DegreeDrop show better performance with intermediate epochs (\ie 20 and 50). 
The conclusion is in line with Fig.~\ref{fig:perform_convergence}, DegreeDrop is capable of converging to better performance with faster training speed.

\subsubsection{Mixing DegreeDrop with DropEdge~\cite{rong2020dropedge}}
We further mix up the two edge dropout mechanisms in sampling the sparsified adjacency matrix $\mathbf{A}_{p}$ during training epochs and evaluate its performance in LayerGCN.
We copy the results of DegreeDrop and DropEdge under each dataset in Table~\ref{tab:full_perform} and compare them with the mixed dropout mechanism in Table~\ref{tab:degree-random}. 
As a result, in most cases, the mixing of DegreeDrop can improve DropEdge's performance, but the performance is still inferior to DegreeDrop. 
These results support the statement that nodes with high degrees are more likely to be over-smoothed~\cite{chen2020simple} and pruning their connections results in better performance.
We further analyze the performance of DegreeDrop with MOOC and Yelp, which are two representative datasets under dense and sparse graph settings. 
\begin{table}[!tbp]	
	\def\arraystretch{1.0}
	\setlength\tabcolsep{6.0pt}
	\centering
	\caption{Performance of LayerGCN with mixed DegreeDrop and DropEdge.}
	 \begin{tabular}{l | l | c c c c }
		\hline
		Datasets & Dropout Types & R@20 & R@50 & N@20 & N@50\\
		\hline %
		\multirow{3}{*}{MOOC} & DropEdge & 0.3571 & 0.5510 & 0.1940 & 0.2421 \\
		& Mixed & 0.3936 & 0.5550 & 0.2062 & 0.2457 \\
		& DegreeDrop & 0.3979 & 0.5572 & 0.2272 & 0.2665 \\
		\hline
		\multirow{3}{*}{Games} & DropEdge & 0.0489 & 0.0882 & 0.0234 & 0.0332 \\
		& Mixed & 0.0489 & 0.0895 & 0.0240 & 0.0339  \\
		& DegreeDrop & 0.0502 & 0.0917 & 0.0254 & 0.0355 \\
		\hline
		\multirow{3}{*}{Food} & DropEdge & 0.0309 & 0.0530 & 0.0171 & 0.0228 \\
		& Mixed & 0.0314 & 0.0546 & 0.0172 & 0.0232 \\
		& DegreeDrop & 0.0318 & 0.0540 & 0.0177 & 0.0234 \\
		\hline
		\multirow{3}{*}{Yelp} & DropEdge & 0.0532 & 0.1017 & 0.0350 & 0.0493 \\
		& Mixed & 0.0519 & 0.1009 & 0.0350 & 0.0493 \\
		& DegreeDrop & 0.0542 & 0.1024 & 0.0360 & 0.0501 \\
		\hline
	\end{tabular}
	\label{tab:degree-random}
\end{table}
\subsubsection{Analysis of DegreeDrop} From Tables~\ref{tab:full_perform} and ~\ref{tab:degree-random}, we observe the performance of DegreeDrop is slightly superior to DropEdge~\cite{rong2020dropedge} yet with faster converge speed. However, on MOOC dataset, we notice DegreeDrop obtains a significant improvement over DropEdge on most evaluation metrics. As a result, we plot the cumulative distribution of degrees (\# of edges) for items in datasets of MOOC and Yelp in Fig.~\ref{fig:cdf_mooc_yelp}. Noting that in Fig.~\ref{fig:cdf_mooc_yelp} we take the square root of node degree ($x$-axis) following the formula of DegreeDrop presented in Eq.~\ref{eq:degreedrop}.
The figure reveals that items in the MOOC dataset have a relatively higher degree than that of Yelp, with 20\% items owning more than 4,00 interactions. Based on~\cite{chen2020simple}, these items are more likely to be over-smoothed in graph learning, as a result, DegreeDrop removes part of their edges with higher probability instead of uniform distribution (DropEdge) and obtains better performance and convergence. 
On the contrary, the distribution of degrees for items in Yelp is highly skewed, roughly 90\% of nodes' degrees are less than 10 when rooted. Hence, the degree probabilities calculated with DegreeDrop are difficult to differentiate these nodes from each other, resulting in a limited improvement in recommendation accuracy over DropEdge.

\subsection{Visualization of Layer Similarities}
Different from LightGCN~\cite{he2020lightgcn} which uses fixed weights or learnable parameters to control the information integration from all hidden layers, LayerGCN adopts a similarity function to dynamically adjust the weight of information from the previous layer. With MOOC datasets, we visualize the similarity values of each layer in Fig.~\ref{fig:layergcn-sim} and compare it with the weights of layers in LightGCN (Fig.~\ref{fig:intro_prob}). 
In Fig.~\ref{fig:layergcn-sim}, we observe that no single layer dominates others in contributing to node representations. Interestingly, we also notice that neighbors from even layers contribute more than their previously odd layers. The reason is that neighbors in even hops have the same type (user/item) as the target node in the bipartite interaction graph. That is in line with our intuition that the same type of nodes should be more similar.

\begin{figure}
	\centering
	\begin{minipage}{.47\linewidth}
		\centering
		\includegraphics[trim=10 10 10 10, clip, width=0.9\linewidth]{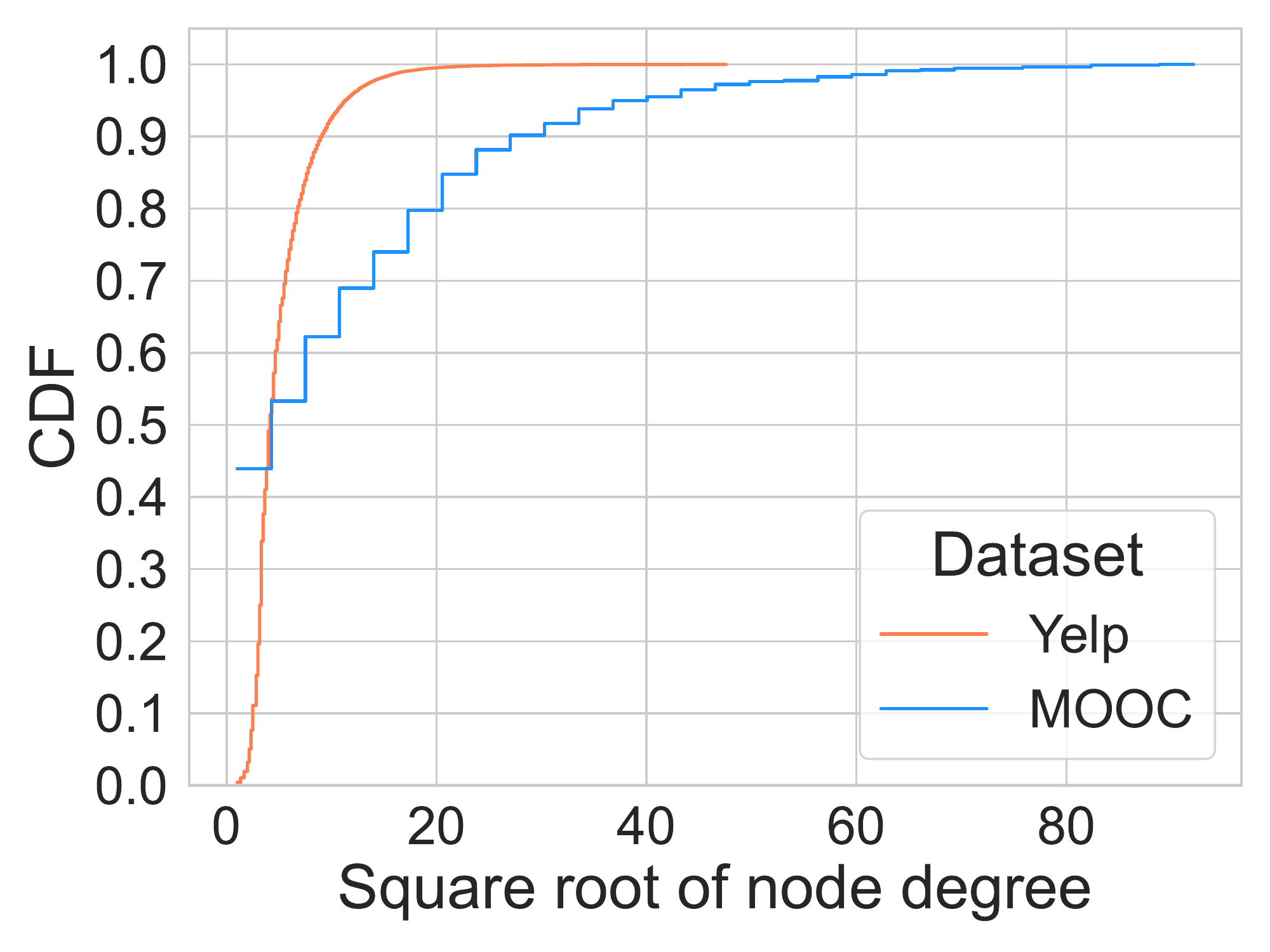}
		\captionsetup{width=.4\linewidth}
		\caption{Distributions of degrees \\ for items in MOOC and Yelp. }
		\label{fig:cdf_mooc_yelp}
	\end{minipage}%
	\begin{minipage}{.47\linewidth}
		\centering
		\includegraphics[trim=10 10 10 10, clip, width=0.92\linewidth]{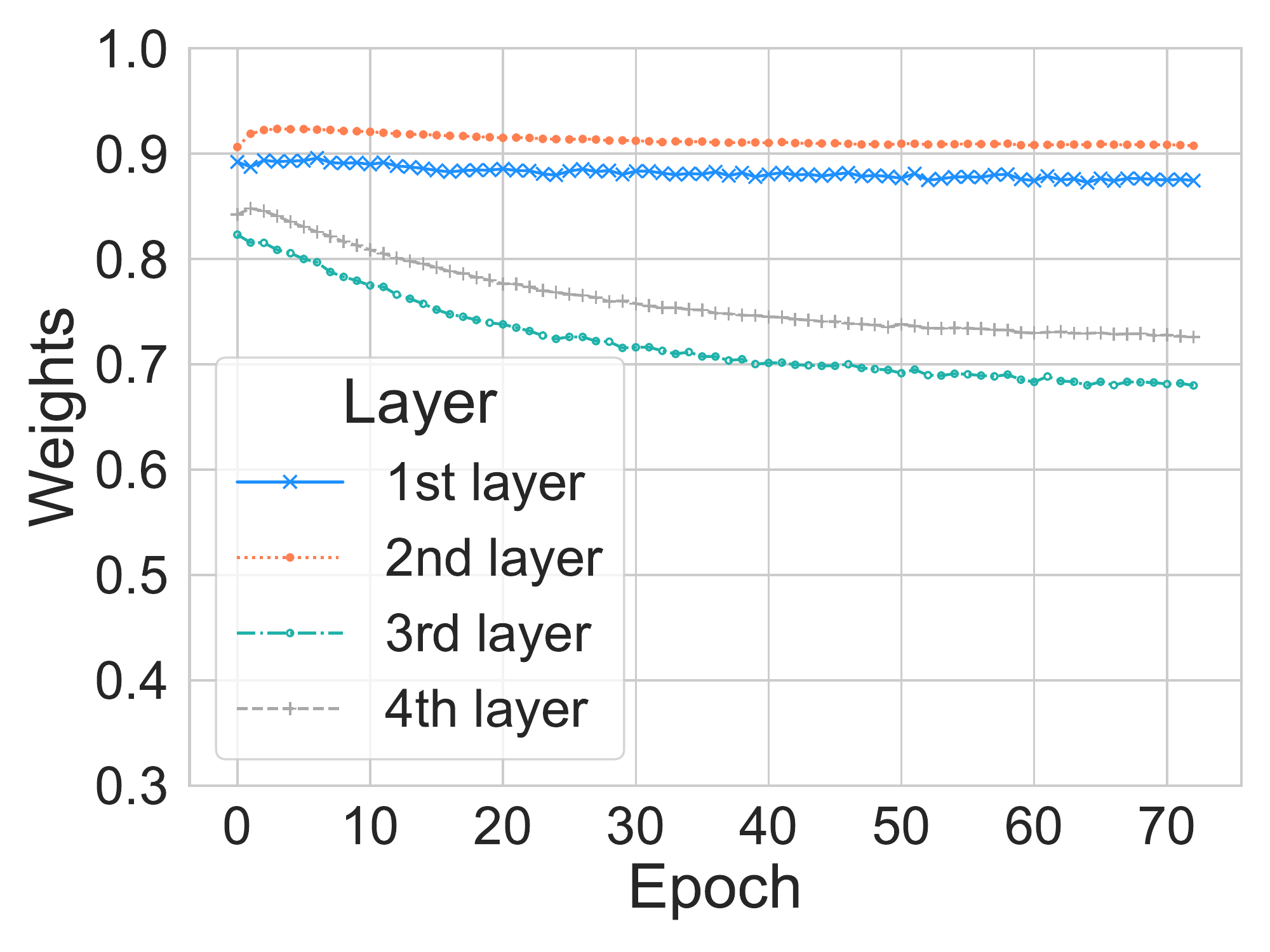}
		\caption{Weights of layers updated during training of LayerGCN. }
		\label{fig:layergcn-sim}
	\end{minipage}
\end{figure}

\subsection{Hyper-parameter Studies}
In this section, we further investigate the performance trends of LayerGCN with respect to (\textit{w.r.t}) different settings of hyper-parameters.
We fix the other parameters under their optimal values with the best performance and relax the studied hyper-parameter under a range of choices to investigate how it affects the performance of LayerGCN.
We analyze each of them as follows.
\subsubsection{Number of Layers $L$}
We search $L$ in the range of [1, 8] with a step size of 1 for LayerGCN and show the results in Fig.~\ref{fig:layer-mooc}.
The best result is observed within 3-layer GCNs for LightGCN, and stacking more layers degrades the recommendation performance.
On the contrary, the proposed LayerGCN shows improved performance over LightGCN when stacking more layers.
The reason is that LayerGCN refines the node representations \wrt to its ego representations.
The layer refinement helps to distill the similar information propagated from the neighboring nodes and discard the irrelevant component.
As a result, the proposed LayerGCN is capable of partially alleviating the over-smoothing issue aforementioned in Section I.

\begin{figure}[bpt]
	\centering
	\includegraphics[trim=10 10 10 10, clip, width=0.46\textwidth]{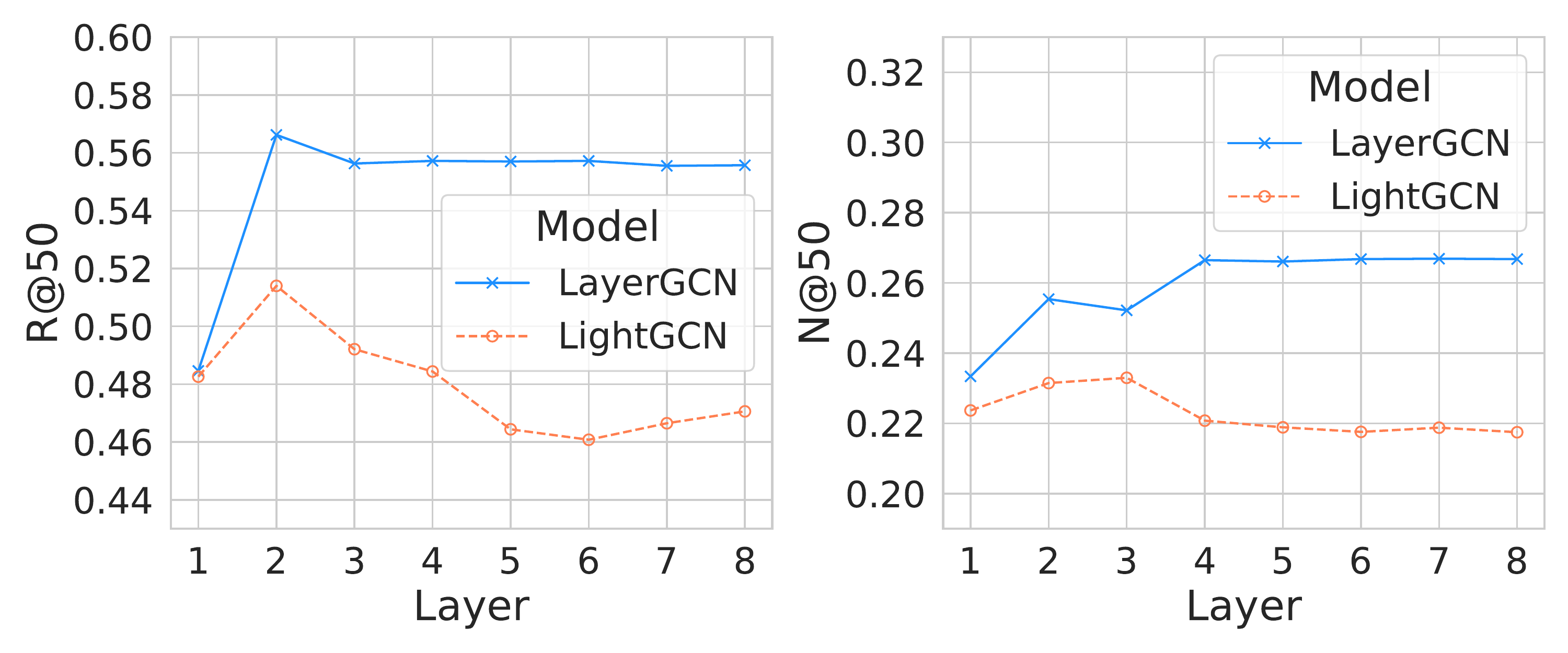}
	\caption{Effect of the number of layers on LayerGCN and LightGCN \wrt the MOOC dataset.}
	\label{fig:layer-mooc}
\end{figure}

\begin{figure}[!h]
	\centering
	\subfloat[Effect of regularization and dropout on MOOC.]{\includegraphics[width=0.85\linewidth]{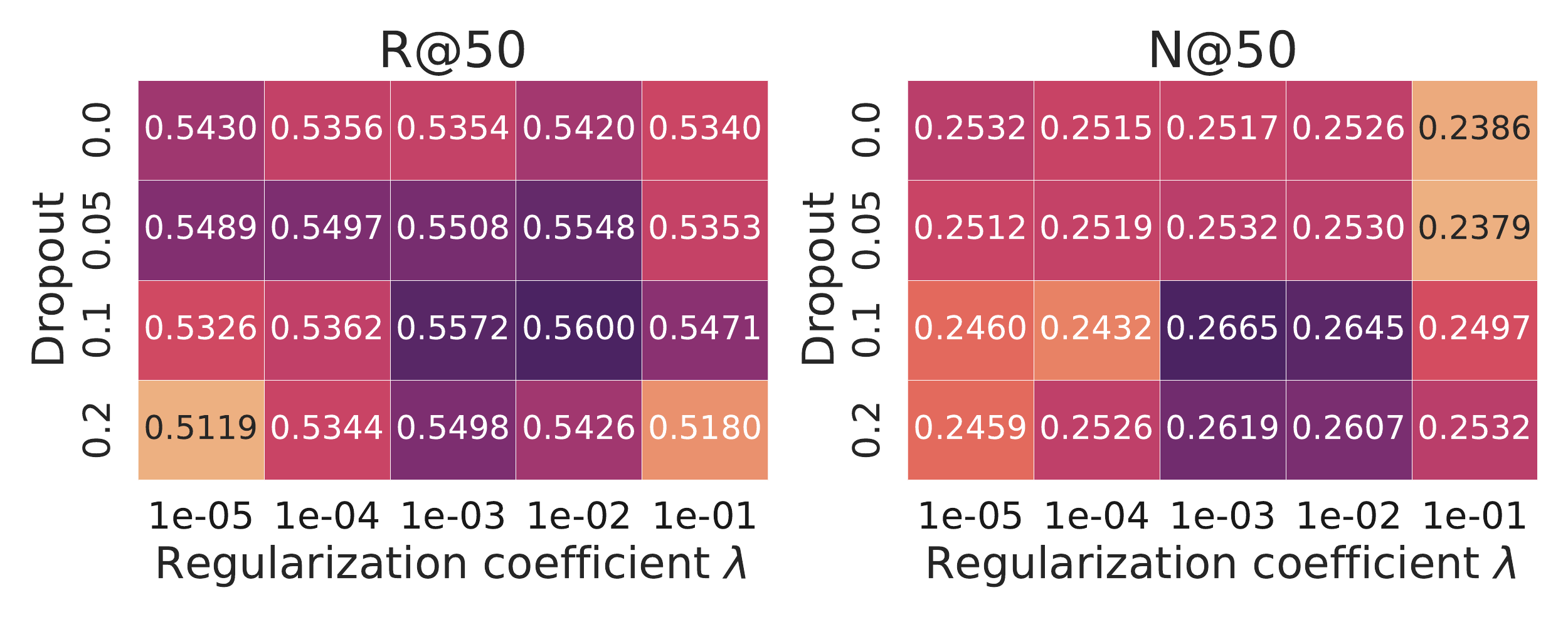}} \hspace{0.1cm}
	\subfloat[Effect of regularization and dropout on Yelp.]{\includegraphics[width=0.85\linewidth]{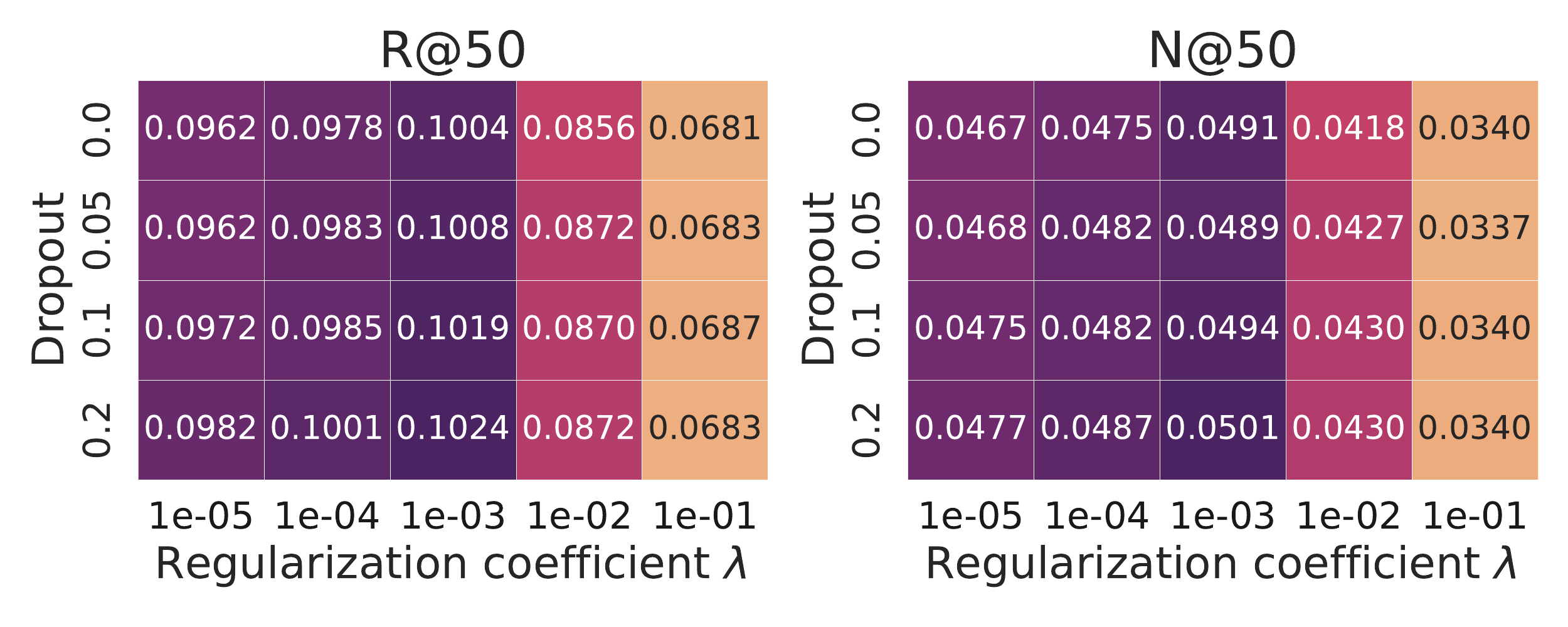}}
	\caption{Performance of LayerGCN \wrt different regularization coefficient $\lambda$ and dropout ratio.  Darker background indicates better recommendation accuracy.}
	\label{fig:hyper_weights}
\end{figure}

\subsubsection{Regularization Coefficient $\lambda$ \textit{vs.} Dropout Ratio}
We search the dropout ratio in \{0.0, 0.05, 0.1, 0.2\}.
Specially, the dropout ratio of 0.0 is the variant of LayerGCN without edge dropout.
The regularization coefficient $\lambda$ is varied in \{$1e-05, 1e-04, \dotsm, 1e-01$\}. 

Fig.~\ref{fig:hyper_weights}(a) shows the edge pruning ratio of 0.1 resulting in best recommendation accuracy on MOOC. Increasing the ratio to 0.2 would degrade the recommendation performance. The potential reason is that edges connecting a hub might be permanently pruned, hence splitting the graph into disconnected subgraphs hampers the information propagation in GCN. 
Compared with Fig.~\ref{fig:hyper_weights}(b), we suggest a low dropout ratio on a dense dataset. 

From both results of Fig.~\ref{fig:hyper_weights}(a) and Fig.~\ref{fig:hyper_weights}(b), we observe LayerGCN is less sensitive to the regularization term.
On both datasets, the optimal value for $\lambda$ is 0.001. A stronger penalty on regularization encourages the model to overcome over-fitting.
However, too strong regularization will negatively affect model training and is not encouraged.

\section{Conclusion and Future Work}
In this work, we first study the over-smoothing problem in existing GCN-based recommendation models.
To tackle the issue, we propose a GCN variant (LayerGCN) that refines the layers during information aggregation.
LayerGCN extracts the information in each propagated layer by dynamically weighting its similarity with the ego layer.
To further alleviate the recommendation degeneration caused by the noise rooted in the user-item interaction graph, we sparsify the graph with a degree-sensitive edge pruning technique.
We theoretically analyze the proposed model can alleviate the over-smoothing issue suffered in current graph-based recommendation models.
Experimental results on four public datasets show our proposed model can significantly improve the baselines with fast training convergence.
In our future work, we would like to study how self-supervised signals~\cite{liu2022graph} can augment the representation learning of LayerGCN.
Also, we will evaluate the performance of LayerGCN under content-based recommendation settings.

\section*{Acknowledgment}

This research is supported, in part, by the Alibaba-NTU Singapore Joint Research Institute, Nanyang Technological University, Singapore, and by JSPS KAKENHI under Grant Number 21H03556.

\bibliographystyle{IEEEtran}
\bibliography{sample.bib}

\begin{thebibliography}{10}
\providecommand{\url}[1]{#1}
\csname url@samestyle\endcsname
\providecommand{\newblock}{\relax}
\providecommand{\bibinfo}[2]{#2}
\providecommand{\BIBentrySTDinterwordspacing}{\spaceskip=0pt\relax}
\providecommand{\BIBentryALTinterwordstretchfactor}{4}
\providecommand{\BIBentryALTinterwordspacing}{\spaceskip=\fontdimen2\font plus
\BIBentryALTinterwordstretchfactor\fontdimen3\font minus
  \fontdimen4\font\relax}
\providecommand{\BIBforeignlanguage}[2]{{%
\expandafter\ifx\csname l@#1\endcsname\relax
\typeout{** WARNING: IEEEtran.bst: No hyphenation pattern has been}%
\typeout{** loaded for the language `#1'. Using the pattern for}%
\typeout{** the default language instead.}%
\else
\language=\csname l@#1\endcsname
\fi
#2}}
\providecommand{\BIBdecl}{\relax}
\BIBdecl

\bibitem{monti2017geometric}
F.~Monti, M.~M. Bronstein, and X.~Bresson, ``Geometric matrix completion with
  recurrent multi-graph neural networks,'' in \emph{Proceedings of the 31st
  International Conference on Neural Information Processing Systems}, ser.
  NIPS'17, 2017, pp. 3700--3710.

\bibitem{wang2019neural}
\BIBentryALTinterwordspacing
X.~Wang, X.~He, M.~Wang, F.~Feng, and T.-S. Chua, ``Neural graph collaborative
  filtering,'' in \emph{Proceedings of the 42nd International ACM SIGIR
  Conference on Research and Development in Information Retrieval}, ser.
  SIGIR'19, 2019, pp. 165--174. [Online]. Available:
  \url{https://doi.org/10.1145/3331184.3331267}
\BIBentrySTDinterwordspacing

\bibitem{he2020lightgcn}
\BIBentryALTinterwordspacing
X.~He, K.~Deng, X.~Wang, Y.~Li, Y.~Zhang, and M.~Wang, ``Lightgcn: Simplifying
  and powering graph convolution network for recommendation,'' in
  \emph{Proceedings of the 43rd International ACM SIGIR Conference on Research
  and Development in Information Retrieval}, ser. SIGIR'20.\hskip 1em plus
  0.5em minus 0.4em\relax New York, NY, USA: Association for Computing
  Machinery, 2020, pp. 639--648. [Online]. Available:
  \url{https://doi.org/10.1145/3397271.3401063}
\BIBentrySTDinterwordspacing

\bibitem{kipf2017semi}
T.~N. Kipf and M.~Welling, ``Semi-supervised classification with graph
  convolutional networks,'' in \emph{International Conference on Learning
  Representations}, 2017.

\bibitem{wu2019simplifying}
F.~Wu, A.~Souza, T.~Zhang, C.~Fifty, T.~Yu, and K.~Weinberger, ``Simplifying
  graph convolutional networks,'' in \emph{International conference on machine
  learning}.\hskip 1em plus 0.5em minus 0.4em\relax PMLR, 2019, pp. 6861--6871.

\bibitem{chen2020revisiting}
L.~Chen, L.~Wu, R.~Hong, K.~Zhang, and M.~Wang, ``Revisiting graph based
  collaborative filtering: A linear residual graph convolutional network
  approach,'' in \emph{Proceedings of the AAAI Conference on Artificial
  Intelligence}, vol.~34, no.~01, 2020, pp. 27--34.

\bibitem{zhang2019hierarchical}
J.~Zhang, B.~Hao, B.~Chen, C.~Li, H.~Chen, and J.~Sun, ``Hierarchical
  reinforcement learning for course recommendation in moocs,'' in
  \emph{Proceedings of the AAAI Conference on Artificial Intelligence},
  vol.~33, no.~01, 2019, pp. 435--442.

\bibitem{yera2016fuzzy}
R.~Yera, J.~Castro, and L.~Mart{\'\i}nez, ``A fuzzy model for managing natural
  noise in recommender systems,'' \emph{Applied Soft Computing}, vol.~40, pp.
  187--198, 2016.

\bibitem{he2016deep}
K.~He, X.~Zhang, S.~Ren, and J.~Sun, ``Deep residual learning for image
  recognition,'' in \emph{Proceedings of the IEEE conference on computer vision
  and pattern recognition}, 2016, pp. 770--778.

\bibitem{chen2020simple}
M.~Chen, Z.~Wei, Z.~Huang, B.~Ding, and Y.~Li, ``Simple and deep graph
  convolutional networks,'' in \emph{International Conference on Machine
  Learning}.\hskip 1em plus 0.5em minus 0.4em\relax PMLR, 2020, pp. 1725--1735.

\bibitem{li2021training}
G.~Li, M.~M{\"u}ller, B.~Ghanem, and V.~Koltun, ``Training graph neural
  networks with 1000 layers,'' in \emph{International Conference on Machine
  Learning}.\hskip 1em plus 0.5em minus 0.4em\relax PMLR, 2021.

\bibitem{rong2020dropedge}
Y.~Rong, W.~Huang, T.~Xu, and J.~Huang, ``Dropedge: Towards deep graph
  convolutional networks on node classification,'' in \emph{International
  Conference on Learning Representations}, 2020.

\bibitem{luo2021learning}
\BIBentryALTinterwordspacing
D.~Luo, W.~Cheng, W.~Yu, B.~Zong, J.~Ni, H.~Chen, and X.~Zhang, ``Learning to
  drop: Robust graph neural network via topological denoising,'' in
  \emph{Proceedings of the 14th ACM International Conference on Web Search and
  Data Mining}, ser. WSDM '21.\hskip 1em plus 0.5em minus 0.4em\relax New York,
  NY, USA: Association for Computing Machinery, 2021, pp. 779--787. [Online].
  Available: \url{https://doi.org/10.1145/3437963.3441734}
\BIBentrySTDinterwordspacing

\bibitem{schlichtkrull2021interpreting}
M.~S. Schlichtkrull, N.~D. Cao, and I.~Titov, ``Interpreting graph neural
  networks for nlp with differentiable edge masking,'' in \emph{International
  Conference on Learning Representations}, 2021.

\bibitem{weisfeiler1968reduction}
B.~Weisfeiler and A.~Leman, ``The reduction of a graph to canonical form and
  the algebra which appears therein,'' \emph{NTI, Series}, vol.~2, no.~9, pp.
  12--16, 1968.

\bibitem{covington2016deep}
P.~Covington, J.~Adams, and E.~Sargin, ``Deep neural networks for youtube
  recommendations,'' in \emph{Proceedings of the 10th ACM conference on
  recommender systems}, 2016, pp. 191--198.

\bibitem{bokde2015matrix}
D.~Bokde, S.~Girase, and D.~Mukhopadhyay, ``Matrix factorization model in
  collaborative filtering algorithms: A survey,'' \emph{Procedia Computer
  Science}, vol.~49, pp. 136--146, 2015.

\bibitem{paterek2007improving}
A.~Paterek, ``Improving regularized singular value decomposition for
  collaborative filtering,'' in \emph{Proceedings of KDD cup and workshop},
  vol. 2007, 2007, pp. 5--8.

\bibitem{wang2015collaborative}
H.~Wang, N.~Wang, and D.-Y. Yeung, ``Collaborative deep learning for
  recommender systems,'' in \emph{Proceedings of the 21th ACM SIGKDD
  international conference on knowledge discovery and data mining}, 2015, pp.
  1235--1244.

\bibitem{he2017neural}
X.~He, L.~Liao, H.~Zhang, L.~Nie, X.~Hu, and T.-S. Chua, ``Neural collaborative
  filtering,'' in \emph{Proceedings of the 26th international conference on
  world wide web}, 2017, pp. 173--182.

\bibitem{zhang2019deep}
S.~Zhang, L.~Yao, A.~Sun, and Y.~Tay, ``Deep learning based recommender system:
  A survey and new perspectives,'' \emph{ACM Computing Surveys (CSUR)},
  vol.~52, no.~1, pp. 1--38, 2019.

\bibitem{berg2018graph}
R.~v.~d. Berg, T.~N. Kipf, and M.~Welling, ``Graph convolutional matrix
  completion,'' in \emph{Proceedings of the 24th ACM SIGKDD international
  conference on knowledge discovery and data mining}, 2018.

\bibitem{ying2018graph}
R.~Ying, R.~He, K.~Chen, P.~Eksombatchai, W.~L. Hamilton, and J.~Leskovec,
  ``Graph convolutional neural networks for web-scale recommender systems,'' in
  \emph{Proceedings of the 24th ACM SIGKDD International Conference on
  Knowledge Discovery \& Data Mining}, 2018, pp. 974--983.

\bibitem{zhang2022diffusion}
L.~Zhang, Y.~Liu, X.~Zhou, C.~Miao, G.~Wang, and H.~Tang, ``Diffusion-based
  graph contrastive learning for recommendation with implicit feedback,'' in
  \emph{International Conference on Database Systems for Advanced
  Applications}.\hskip 1em plus 0.5em minus 0.4em\relax Springer, 2022, pp.
  232--247.

\bibitem{zhou2021selfcf}
X.~Zhou, A.~Sun, Y.~Liu, J.~Zhang, and C.~Miao, ``Selfcf: A simple framework
  for self-supervised collaborative filtering,'' 2021.

\bibitem{lee2021bootstrapping}
D.~Lee, S.~Kang, H.~Ju, C.~Park, and H.~Yu, ``Bootstrapping user and item
  representations for one-class collaborative filtering,'' in \emph{Proceedings
  of the 44th International ACM SIGIR Conference on Research and Development in
  Information Retrieval}, 2021.

\bibitem{mao2021ultragcn}
K.~Mao, J.~Zhu, X.~Xiao, B.~Lu, Z.~Wang, and X.~He, ``Ultragcn: ultra
  simplification of graph convolutional networks for recommendation,'' in
  \emph{Proceedings of the 30th ACM International Conference on Information \&
  Knowledge Management}, 2021, pp. 1253--1262.

\bibitem{liu2021interest}
F.~Liu, Z.~Cheng, L.~Zhu, Z.~Gao, and L.~Nie, ``Interest-aware message-passing
  gcn for recommendation,'' in \emph{Proceedings of the Web Conference 2021},
  2021, pp. 1296--1305.

\bibitem{he2016vbpr}
R.~He and J.~McAuley, ``Vbpr: visual bayesian personalized ranking from
  implicit feedback,'' in \emph{Proceedings of the AAAI conference on
  artificial intelligence}, vol.~30, no.~1, 2016.

\bibitem{rendle2009bpr}
S.~Rendle, C.~Freudenthaler, Z.~Gantner, and L.~Schmidt-Thieme, ``Bpr: Bayesian
  personalized ranking from implicit feedback,'' in \emph{Proceedings of the
  Twenty-Fifth Conference on Uncertainty in Artificial Intelligence}, 2009, pp.
  452--461.

\bibitem{wang2021learning}
X.~Wang, T.~Huang, D.~Wang, Y.~Yuan, Z.~Liu, X.~He, and T.-S. Chua, ``Learning
  intents behind interactions with knowledge graph for recommendation,'' in
  \emph{Proceedings of the Web Conference 2021}, 2021, pp. 878--887.

\bibitem{lops2011content}
P.~Lops, M.~d. Gemmis, and G.~Semeraro, ``Content-based recommender systems:
  State of the art and trends,'' \emph{Recommender systems handbook}, pp.
  73--105, 2011.

\bibitem{guo2020survey}
Q.~Guo, F.~Zhuang, C.~Qin, H.~Zhu, X.~Xie, H.~Xiong, and Q.~He, ``A survey on
  knowledge graph-based recommender systems,'' \emph{IEEE Transactions on
  Knowledge and Data Engineering}, 2020.

\bibitem{zhang2021mining}
J.~Zhang, Y.~Zhu, Q.~Liu, S.~Wu, S.~Wang, and L.~Wang, ``Mining latent
  structures for multimedia recommendation,'' in \emph{Proceedings of the 29th
  ACM International Conference on Multimedia}, 2021, pp. 3872--3880.

\bibitem{zhou2022bootstrap}
X.~Zhou, H.~Zhou, Y.~Liu, Z.~Zeng, C.~Miao, P.~Wang, Y.~You, and F.~Jiang,
  ``Bootstrap latent representations for multi-modal recommendation,''
  \emph{arXiv preprint arXiv:2207.05969}, 2022.

\bibitem{zhou2022tale}
X.~Zhou, ``A tale of two graphs: Freezing and denoising graph structures for
  multimodal recommendation,'' \emph{arXiv preprint arXiv:2211.06924}, 2022.

\bibitem{velickovic2019deep}
P.~Velickovic, W.~Fedus, W.~L. Hamilton, P.~Li{\`o}, Y.~Bengio, and R.~D.
  Hjelm, ``Deep graph infomax,'' in \emph{International Conference on Learning
  Representations (Poster)}, 2019.

\bibitem{louizos2018learning}
C.~Louizos, M.~Welling, and D.~P. Kingma, ``Learning sparse neural networks
  through $l\_0$ regularization,'' in \emph{International Conference on
  Learning Representations}, 2018.

\bibitem{li2018deeper}
Q.~Li, Z.~Han, and X.-M. Wu, ``Deeper insights into graph convolutional
  networks for semi-supervised learning,'' in \emph{Proceedings of the AAAI
  Conference on Artificial Intelligence}, vol.~32, no.~1, 2018.

\bibitem{oono2020graph}
K.~Oono and T.~Suzuki, ``Graph neural networks exponentially lose expressive
  power for node classification,'' in \emph{International Conference on
  Learning Representations}, 2020.

\bibitem{sundararaman2020methods}
D.~Sundararaman, S.~Si, V.~Subramanian, G.~Wang, D.~Hazarika, and L.~Carin,
  ``Methods for numeracy-preserving word embeddings,'' in \emph{Proceedings of
  the 2020 Conference on Empirical Methods in Natural Language Processing
  (EMNLP)}, 2020, pp. 4742--4753.

\bibitem{xu2019powerful}
K.~Xu, W.~Hu, J.~Leskovec, and S.~Jegelka, ``How powerful are graph neural
  networks?'' in \emph{International Conference on Learning Representations},
  2019.

\bibitem{kumar2017weight}
S.~K. Kumar, ``On weight initialization in deep neural networks,'' \emph{arXiv
  preprint arXiv:1704.08863}, 2017.

\bibitem{velivckovic2017graph}
P.~Veli{\v{c}}kovi{\'c}, G.~Cucurull, A.~Casanova, A.~Romero, P.~Lio, and
  Y.~Bengio, ``Graph attention networks,'' in \emph{International Conference on
  Learning Representations}, 2018.

\bibitem{brody2022how}
S.~Brody, U.~Alon, and E.~Yahav, ``How attentive are graph attention
  networks?'' in \emph{International Conference on Learning Representations},
  2022.

\bibitem{ji2020critical}
Y.~Ji, A.~Sun, J.~Zhang, and C.~Li, ``A critical study on data leakage in
  recommender system offline evaluation,'' \emph{arXiv preprint
  arXiv:2010.11060}, 2020.

\bibitem{ni2019justifying}
J.~Ni, J.~Li, and J.~McAuley, ``Justifying recommendations using
  distantly-labeled reviews and fine-grained aspects,'' in \emph{Proceedings of
  the 2019 Conference on Empirical Methods in Natural Language Processing and
  the 9th International Joint Conference on Natural Language Processing
  (EMNLP-IJCNLP)}, 2019, pp. 188--197.

\bibitem{he2016fast}
X.~He, H.~Zhang, M.-Y. Kan, and T.-S. Chua, ``Fast matrix factorization for
  online recommendation with implicit feedback,'' in \emph{Proceedings of the
  39th International ACM SIGIR conference on Research and Development in
  Information Retrieval}, 2016, pp. 549--558.

\bibitem{tan2020learning}
Q.~Tan, N.~Liu, X.~Zhao, H.~Yang, J.~Zhou, and X.~Hu, ``Learning to hash with
  graph neural networks for recommender systems,'' in \emph{Proceedings of The
  Web Conference 2020}, 2020, pp. 1988--1998.

\bibitem{liang2018variational}
D.~Liang, R.~G. Krishnan, M.~D. Hoffman, and T.~Jebara, ``Variational
  autoencoders for collaborative filtering,'' in \emph{Proceedings of the 2018
  world wide web conference}, 2018, pp. 689--698.

\bibitem{chen2020efficient}
C.~Chen, M.~Zhang, Y.~Zhang, W.~Ma, Y.~Liu, and S.~Ma, ``Efficient
  heterogeneous collaborative filtering without negative sampling for
  recommendation,'' in \emph{Proceedings of the AAAI Conference on Artificial
  Intelligence}, vol.~34, no.~01, 2020, pp. 19--26.

\bibitem{glorot2010understanding}
X.~Glorot and Y.~Bengio, ``Understanding the difficulty of training deep
  feedforward neural networks,'' in \emph{Proceedings of the thirteenth
  international conference on artificial intelligence and statistics}.\hskip
  1em plus 0.5em minus 0.4em\relax JMLR Workshop and Conference Proceedings,
  2010, pp. 249--256.

\bibitem{kingma2015adam}
D.~P. Kingma and J.~Ba, ``Adam: A method for stochastic optimization,'' in
  \emph{International Conference on Learning Representations}, 2015.

\bibitem{NEURIPS2019_bdbca288}
A.~Paszke, S.~Gross, F.~Massa, A.~Lerer, J.~Bradbury, G.~Chanan, T.~Killeen,
  Z.~Lin, N.~Gimelshein, L.~Antiga, A.~Desmaison, A.~Kopf, E.~Yang, Z.~DeVito,
  M.~Raison, A.~Tejani, S.~Chilamkurthy, B.~Steiner, L.~Fang, J.~Bai, and
  S.~Chintala, ``Pytorch: An imperative style, high-performance deep learning
  library,'' in \emph{Advances in Neural Information Processing Systems},
  vol.~32.\hskip 1em plus 0.5em minus 0.4em\relax Curran Associates, Inc.,
  2019.

\bibitem{liu2022graph}
Y.~Liu, M.~Jin, S.~Pan, C.~Zhou, Y.~Zheng, F.~Xia, and P.~Yu, ``Graph
  self-supervised learning: A survey,'' \emph{IEEE Transactions on Knowledge
  and Data Engineering}, 2022.

\end{thebibliography}

\end{document}